\documentclass[12pt,showpacs,aps,nofootinbib]{revtex4}
\usepackage{amsmath,amssymb,graphicx,epsfig,color,ulem,cancel,xcolor}
\def\comment#1{}

\def\nn{\nonumber}
\def\ds{\mathrm{dS}}
\def\dsd{\mathrm{dS_{D}}}
\def\x{\mathbf{x}}
\def\k{\mathbf{k}}
\def\lambdam{\lambda_{\mathrm{m}}}
\def\mds{m_{\mathrm{ds}}}
\def\rnd{\partial}
\def\wwp{\mathrm{W}_{\kappa,\gamma}}
\def\wwm{\mathrm{W}_{\kappa,-\gamma}}
\def\wmp{\mathrm{M}_{\kappa,\gamma}}
\def\wmm{\mathrm{M}_{\kappa,-\gamma}}
\def\w{\mathrm{W}}
\def\m{\mathrm{M}}

\def\I{\mathrm{I}}
\def\sem{\mathrm{sem}}
\begin{document}
%%%%%%%%%%%%%%%%%%%%%%%%%%%%%%%%%%%%%%%%%%%%%%%%%%%%%%%%%%%%%%%%%%%%%%%%%%%%%%%%%%%%%%%%%%%%%%%%%%%%%%%%%%%%%%%%%%%%%%%%%%%%%%%%%%%%%%%%%%%%%%%%%%%%%%%%%%%%%%%
%%%%%%%%%%%%%%%%%%%%%%%%%%%%%%%%%%%%%%%%%%%%%%%%%%%%%%%%%%%%%%%%%%%%%%%%%%%%%%%%%%%%%%%%%%%%%%%%%%%%%%%%%%%%%%%%%%%%%%%%%%%%%%%%%%%%%%%%%%%%%%%%%%%%%%%%%%%%%%%
\title{Scalar current of created pairs by Schwinger mechanism in de~Sitter spacetime}
\author{Ehsan~Bavarsad$^{1}$}\email{bavarsad@kashanu.ac.ir}
\author{Cl\'{e}ment~Stahl$^{2,3,4,5}$}\email{clement.stahl@icranet.org}
\author{She-Sheng~Xue$^{2,3}$}\email{xue@icra.it}
\affiliation{$^{1}$Department of Physics, University of Kashan, 8731753153 Kashan, Iran \\
$^{2}$ICRANet, Piazzale della Repubblica 10 65122 Pescara, Italy \\
$^{3}$Dipartimento di Fisica Universit\`a di Roma "La Sapienza," Piazzale Aldo Moro 5, 00185 Rome, Italy \\
$^{4}$Universit\'e de Nice Sophia Antipolis, 28 Avenue de Valrose, 06103 Nice Cedex 2, France \\
$^{5}$Nordita KTH Royal Institute of Technology and Stockholm University, Roslagstullsbacken 23, 114 21 Stockholm, Sweden}
%\date{\tt{Revised Version: \today}}
%%%%%%%%%%%%%%%%%%%%%%%%%%%%%%%%%%%%%%%%%%%%%%%%%%%%%%%%%%%%%%%%%%%%%%%%%%%%%%%%%%%%%%%%%%%%%%%%%%%%%%%%%%%%%%%%%%%%%%%%%%%%%%%%%%%%%%%%%%%%%%%%%%%%%%%%%%%%%%%
%%%%%%%%%%%%%%%%%%%%%%%%%%%%%%%%%%%%%%%%%%%%%%%%%%%%%%%%%%%%%%%%%%%%%%%%%%%%%%%%%%%%%%%%%%%%%%%%%%%%%%%%%%%%%%%%%%%%%%%%%%%%%%%%%%%%%%%%%%%%%%%%%%%%%%%%%%%%%%%
\begin{abstract}
We consider a charged scalar field in a $D$-dimensional de~Sitter spacetime and investigate pair creation by a Schwinger mechanism in a constant
electric field background.
Using a semiclassical approximation the current of the created pairs has been estimated.
We find that the semiclassical current of the created pairs in the strong electric field limit responds as $E^{\frac{D}{2}}$.
Going further but restricting to $D=3$ dimensional de~Sitter spacetime, the quantum expectation value of the spacelike component of the induced
current has been computed in the in-vacuum state by applying an adiabatic subtraction scheme.
We find that, in the strong electric field limit, the current responds as $E^{\frac{3}{2}}$.
In the weak electric field limit the current has a linear response in $E$ and an inverse dependence on the mass of the scalar field.
In the case of a massless scalar field, the current varies with $E^{-1}$ which leads to a phenomenon of infrared hyperconductivity.
A new relation between infrared hyperconductivity, tachyons, and conformality is discussed, and a scheme to avoid an infrared hyperconductivity
regime is proposed.
In $D$ dimension, we eventually presented some first estimates of the backreaction of the Schwinger pairs to the gravitational field, and we find
a decrease of the Hubble constant due to the pair creation.
\end{abstract}
\maketitle
%\tableofcontents
%%%%%%%%%%%%%%%%%%%%%%%%%%%%%%%%%%%%%%%%%%%%%%%%%%%%%%%%%%%%%%%%%%%%%%%%%%%%%%%%%%%%%%%%%%%%%%%%%%%%%%%%%%%%%%%%%%%%%%%%%%%%%%%%%%%%%%%%%%%%%%%%%%%%%%%%%%%%%%%
%%%%%%%%%%%%%%%%%%%%%%%%%%%%%%%%%%%%%%%%%%%%%%%%%%%%%%%%%%%%%%%%%%%%%%%%%%%%%%%%%%%%%%%%%%%%%%%%%%%%%%%%%%%%%%%%%%%%%%%%%%%%%%%%%%%%%%%%%%%%%%%%%%%%%%%%%%%%%%%
\section{\label{sec:intro}Introduction}
The purpose of this paper is to look at Schwinger pair creation in $D=1+d$ dimensional de~Sitter spacetime ($\dsd$), with special emphasis on the
case of $D=3$ dimension.
The Schwinger effect, i.e., pair production by a strong electric field, is a nonperturbative effect of quantum field theory in flat spacetime which
was initially discovered in the pioneers' works \cite{Schwinger:1951nm}; for a recent review see, e.g., \cite{Gelis:2015kya}.
Despite tremendous efforts on the experimental point of view, it has never been detected so far.
The main reason is that the number of pairs created is exponentially damped before a critical value for the electric field $E_{\mathrm{critical}}\simeq10^{18}V/m$ \cite{DiPiazza:2011tq}.
New laser facilities \cite{Laser1,Laser2,Laser3,Laser4} are planned to be operational in the next ten years and might approach this critical
electric field.
In the meantime, another proposal is to change the perspective: whereas all the experiments aiming at detecting the Schwinger effect are conducted
on Earth, one could look for the Schwinger effect in astrophysical and cosmological systems.
The review \cite{Ruffini:2009hg} describes some of these ideas, and the stress is on the backreaction of the created pairs and its application to
astrophysics.
During inflationary magnetogenesis strong electric fields are also produced \cite{Durrer:2013pga}, which provides motivation for considering
Schwinger pair production in this context.
Furthermore, pairs can also be created by gravitational fields, e.g., in dS \cite{Mottola:1984ar} and quasi-dS \cite{Baumann:2014nda}, which is
sometimes referred to as the cosmological Schwinger effect \cite{Martin:2007bw}.
These effects can also be essential for the interaction and balance between matter-field and cosmological constant in the Universe evolution \cite{Xue:2014kna,Xue:2015tmw}.
\par
Recently, the combination of electrical and gravitational pair creation was studied in depth for various types of particles and spacetime dimension
\cite{Frob:2014zka,Kobayashi:2014zza,Fischler:2014ama,Stahl:2015gaa,Hayashinaka:2016qqn,Hayashinaka:2016dnt,Yokoyama:2015wws}.
In \cite{Frob:2014zka} and \cite{Kobayashi:2014zza}, the authors computed the Schwinger effect for a charged scalar test field in $\ds_{2}$ and
$\ds_{4}$, respectively.
In \cite{Stahl:2015gaa} and \cite{Hayashinaka:2016qqn}, the generalization to $\ds_{2}$ and $\ds_{4}$, respectively, for fermionic particles was
performed aiming at checking if the known equivalence in flat spacetime between boson and fermion for a constant electric field still holds.
The answer was that there was a difference between boson and fermion.
To see that, it was necessary to compute the induced current which turns out, as also noted in \cite{Frob:2014zka,Kobayashi:2014zza}, to be the
right quantity to describe the Schwinger effect in curved spacetime.
Indeed, it is not plagued by the need of the notion of particle in the adiabatic future which allows one to explore a broader parameter space.
\par
But to cure infinities arising from momentum integration, this current needs to be renormalized.
The adiabatic subtraction is the most used method.
The Pauli-Villars method was implemented in \cite{Frob:2014zka} and can be shown to agree with the adiabatic subtraction.\footnote{Private
communication between Eckhard Strobel and Cl\'ement Stahl.}
In \cite{Hayashinaka:2016dnt}, the point-splitting method was shown to agree with the adiabatic subtraction in $\ds_{4}$ for the boson.
In \cite{Kobayashi:2014zza,Stahl:2015gaa}, an adiabatic subtraction method was used to regularize the current.
To further explore the validity of the use of the adiabatic subtraction, it is legitimate to look at $\ds_{3}$ and see which kinds of divergences
arise and how they are cured.
\par
In \cite{Dabrowski:2014ica}, the Schwinger effect in $\ds_{3}$ has been investigated as an example of odd dimensional dS, and it has been shown
that no particle production occurs in odd dimensional global dS.
However, in our study on the Poincar\'e patch together with an electric field, we report particle production.
One more motivation to look in depth at an odd dimensional quantum field theory is that those theories sometimes exhibit strange behaviors.
For instance, dimensional regularization shows no one-loop ultraviolet divergences because it only registers logarithmic divergences and all the
divergences are power law in an odd dimension.
For $D=3$ in curved spacetime with an electric field, we will, however, report a linear divergence arising from the electromagnetic side of the
theory.
Regarding infrared phenomena, we will also report a different behavior than in an even dimension which is also the case in flat spacetime.
Our work aims at completing the picture of the Schwinger effect in $\ds_{3}$ by computing the induced current.
\par
All the works so far describing pair creation in $\ds$ under the influence of a strong electric field \cite{Frob:2014zka,Kobayashi:2014zza,Fischler:2014ama,Stahl:2015gaa,Hayashinaka:2016qqn,Hayashinaka:2016dnt,Yokoyama:2015wws}
assumed two backgrounds, i.e., electromagnetic and gravitational.
In \cite{Kobayashi:2014zza} and \cite{Stahl:2016geq}, the backreaction effects due to the created Schwinger pairs on the electromagnetic field
background have been investigated, while in this case, there is no report on the backreaction to the gravity sector.
In order to investigate the backreaction to the gravitational background one needs the energy-momentum tensor of the created particles.
In the absence of the electromagnetic background, i.e., in the presence of a purely gravitational field, the energy-momentum tensor of the created
scalar \cite{Mottola:1984ar,Parker:1974qw,Fulling:1974zr,Dowker:1975tf,Habib:1999cs,LopezNacir:2007fdi,Markkanen:2016aes} and fermion
\cite{Landete:2013lpa} particles in a spatially flat Friedmann-Lemaitre-Robertson-Walker universe (considering $\ds_{4}$ as a special case) have
been computed.
In \cite{Mottola:1984ar}, the author examined backreaction to the gravitational field and found that the particle creation leads to a decrease of
the cosmological constant, whereas in \cite{Parker:1974qw,Fulling:1974zr,Dowker:1975tf,Landete:2013lpa}, the authors mainly developed the
renormalization theory.
In the last part of this article, we will present some preliminary results on backreaction to the gravitational field, namely, an estimate of the
variation of the Hubble constant due to the influence of the gravitational field.
We hope to come back to this issue in the future by solving fully the Einstein equation.
\par
The article is organized as follows: in Sec.~\ref{sec:dS}, solutions of the Klein-Gordon equation are obtained.
The pair creation is examined in Sec.~\ref{sec:creat}.
In Sec.~\ref{sec:curent} the quantum vacuum expectation value of induced current is investigated.
Section~\ref{sec:IRHC} is devoted to a discussion about the phenomenon of infrared hyperconductivity.
In Sec.~\ref{sec:gravity}, we present some first results on backreaction of the created pairs in $D$ dimension to the gravity sector of theory.
Eventually, in Sec.~\ref{sec:concl}, we give some concluding remarks.
For the sake of clarity, we relegated some of the technical calculations to the appendixes.
In Appendix~\ref{app:sem}, an alternative method to derive the main result of Sec.~\ref{sec:creat} is given.
In Appendix~\ref{app:math}, some useful properties of mathematical functions are represented.
In Appendix~\ref{app:int} the computation of the current integral is reviewed.
%%%%%%%%%%%%%%%%%%%%%%%%%%%%%%%%%%%%%%%%%%%%%%%%%%%%%%%%%%%%%%%%%%%%%%%%%%%%%%%%%%%%%%%%%%%%%%%%%%%%%%%%%%%%%%%%%%%%%%%%%%%%%%%%%%%%%%%%%%%%%%%%%%%%%%%%%%%%%%%
%%%%%%%%%%%%%%%%%%%%%%%%%%%%%%%%%%%%%%%%%%%%%%%%%%%%%%%%%%%%%%%%%%%%%%%%%%%%%%%%%%%%%%%%%%%%%%%%%%%%%%%%%%%%%%%%%%%%%%%%%%%%%%%%%%%%%%%%%%%%%%%%%%%%%%%%%%%%%%%
\section{\label{sec:dS}Solutions of the Klein-Gordon equation}
To study the vacuum expectation value of the current operator of a charged scalar field coupled to a constant electric field background in a
$\dsd$, the field operator is needed.
The field operator contains mode functions which are obtained by solving the Klein-Gordon equation. The $\dsd$ metric can be read from the line
element in the half of $\dsd$ manifold
\begin{align}\label{line}
ds^{2}=dt^{2}-e^{2Ht}d\x^{2}, &&t\in(-\infty,+\infty), &&\x\in\mathbb{R}^{d}.
\end{align}
It corresponds to the line element of the flat Robertson-Walker universe with its scale factor $a(t)=\exp(Ht)$, $t$ is the proper time, and $H$ is
the Hubble constant. In terms of the conformal time
\begin{align}\label{time}
\tau&=-\frac{1}{H}e^{-Ht}, & \tau&\in(-\infty,0),
\end{align}
the line element~(\ref{line}) takes the form
\begin{equation}\label{metric}
ds^{2}=\Omega^{2}(\tau)\big(d\tau^{2}-d\x^{2}\big),
\end{equation}
where
\begin{equation}\label{Omega}
\Omega(\tau)=-\frac{1}{\tau H},
\end{equation}
revealing that this portion of dS is conformal to a portion of Minkowski spacetime.
%%%%%%%%%%%%%%%%%%%%%%%%%%%%%%%%%%%%%%%%%%%%%%%%%%%%%%%%%%%%%%%%%%%%%%%%%%%%%%%%%%%%%%%%%%%%%%%%%%%%%%%%%%%%%%%%%%%%%%%%%%%%%%%%%%%%%%%%%%%%%%%%%%%%%%%%%%%%%%%
%%%%%%%%%%%%%%%%%%%%%%%%%%%%%%%%%%%%%%%%%%%%%%%%%%%%%%%%%%%%%%%%%%%%%%%%%%%%%%%%%%%%%%%%%%%%%%%%%%%%%%%%%%%%%%%%%%%%%%%%%%%%%%%%%%%%%%%%%%%%%%%%%%%%%%%%%%%%%%%
\subsection{\label{sec:KG}The Klein-Gordon equation}
In order to obtain solutions of the Klein-Gordon equation in the presence of a constant electric field background in $\dsd$, we consider the action
of scalar QED,
\begin{equation}\label{action}
S=\int{d^D}x\sqrt{|g|}\Big\{g^{\mu\nu}\big(\rnd_{\mu}+ieA_{\mu}\big)\varphi\big(\rnd_{\nu}-ieA_{\nu}\big)
\varphi^{\ast}-\big(m^{2}+\xi R\big)\varphi\varphi^{\ast}-\frac{1}{4}F_{\mu\nu}F^{\mu\nu}\Big\},
\end{equation}
where $\varphi(x)$ is a complex scalar field with mass $m$ and electrical charge $e$. The $\dsd$ metric $g_{\mu\nu}$ reads from Eq.~(\ref{metric}),
$|g|$ is the absolute value of its determinant, $R$ is the scalar curvature, and $\xi$ is the dimensionless conformal coupling.
The introduction of the conformal coupling is done to make the theory more general and arises naturally in string inflation framework \cite{Stahl:2016qjs,Kachru:2003sx}. The vector potential describing a constant electric field background is
\begin{equation}\label{vector}
A_{\mu}(\tau)=-\frac{E}{H^{2}\tau}\delta_{\mu1},
\end{equation}
where $E$ is constant. Our convention for indices is that we label the spatial dimension with arabic-persian numerals, e.g.,~$1,2,\ldots$, and
that letters, e.g.,~$x,y,\ldots$, are reserved for Fourier space.
Then, the only nonzero components of the electromagnetic field strength tensor are
\begin{align}\label{tensor}
F_{01}&=-F_{10} \nn\\
&=\rnd_{0}A_{1}-\rnd_{1}A_{0}=\Omega^{2}(\tau)E.
\end{align}
We derive the equation of motion for the scalar field $\varphi$ by using the Euler-Lagrange equation for the Lagrangian coming from the
action~(\ref{action}). We then obtain the Klein-Gordon equation
\begin{equation}\label{kgeq}
\frac{1}{\sqrt{g}}\rnd_{\mu}\big(\sqrt{g}g^{\mu\nu}\rnd_{\nu}\varphi\big)+2ieg^{\mu\nu}A_{\nu}\rnd_{\mu}\varphi
-e^{2}g^{\mu\nu}A_{\mu}A_{\nu}\varphi+\mds^{2}\varphi=0,
\end{equation}
where we defined
\begin{equation}\label{mds}
\mds^{2}:=m^{2}+\xi R.
\end{equation}
After substituting explicit expressions of the $\dsd$ metric and the vector potential given in Eqs.~(\ref{metric}) and~(\ref{vector}), respectively, Eq.~(\ref{kgeq}) takes the form
\begin{equation}\label{phieq}
\bigg[\rnd_{0}^{2}-\delta^{ij}\rnd_{i}\rnd_{j}+(D-2)H\Omega(\tau)\rnd_{0}-\frac{2ieE}{H}\Omega(\tau)\rnd_{1}
+\Big(\frac{e^{2}E^{2}}{H^{2}}+\mds^{2}\Big)\Omega^{2}(\tau)\bigg]\varphi(x)=0.
\end{equation}
If we define
\begin{equation}\label{varphi}
\tilde{\varphi}(x):=\Omega^{\frac{D-2}{2}}(\tau)\varphi(x),
\end{equation}
it can be shown that Eq.~(\ref{phieq}) leads to
\begin{equation}\label{vphieq}
\bigg[\rnd_{0}^{2}-\delta^{ij}\rnd_{i}\rnd_{j}+\frac{2ieE}{\tau H^{2}}\rnd_{1}+\frac{1}{\tau^{2}}
\Big(\frac{e^{2}E^{2}}{H^{4}}+\frac{\mds^{2}}{H^{2}}
+\frac{1-d^{2}}{4}\Big)\bigg]\tilde{\varphi}(x)=0.
\end{equation}
Based on the invariance of Eq.~(\ref{vphieq}) under translations along the spatial directions, let
\begin{equation}\label{fpm}
\tilde{\varphi}(\tau,\x)=e^{\pm i\k\cdot\x}f^{\pm}(\tau),
\end{equation}
where the superscript $\pm$ denotes the positive and negative frequency solutions, respectively.
Substituting~(\ref{fpm}) into Eq.~(\ref{vphieq}) leads to
\begin{equation}\label{feq}
\frac{d^{2}}{dz_{\pm}^{2}}f^{\pm}(z_{\pm})+\Big(-\frac{1}{4}+\frac{\kappa}{z_{\pm}}
+\frac{1/4-\gamma^{2}}{z_{\pm}^{2}}\Big)f^{\pm}(z_{\pm})=0,
\end{equation}
where the variables $z_{+}$ and $z_{-}$ are defined by
\begin{align}\label{zpm}
z_{+}&:=+2ik\tau, & z_{-}&:=e^{i\pi}z_{+}=-2ik\tau,
\end{align}
with $k=|\k|$. In terms of dimensionless parameters
\begin{align}\label{lambda}
\lambdam&:=\frac{\mds}{H}, &
\lambda&:=-\frac{eE}{H^{2}}, &
\rho&:=+\big(\lambdam^{2}+\lambda^{2}\big)^{\frac{1}{2}}, &
r&:=\frac{k_{x}}{k},
\end{align}
the coefficients $\kappa$ and $\gamma$ read
\begin{eqnarray}
\kappa&=&-i\lambda r, \label{kappa} \\
\gamma^{2}&=&\frac{d^{2}}{4}-\lambdam^{2}-\lambda^{2}. \label{gamma}
\end{eqnarray}
In Secs.~\ref{sec:dS} and~\ref{sec:creat} of this article we consider that $\gamma^{2}<0$, and then the coefficient $\gamma$ becomes purely
imaginary; in this case, we use the convention $\gamma=+i|\gamma|$.
Equation~(\ref{feq}) is the Whittaker differential equation, and its most general solution in terms of the conventions of \cite{book:Nist} can be
written as
\begin{equation}\label{general}
f^{\pm}(z_{\pm})=C_{1}\w_{\kappa,\pm\gamma}(z_{\pm})+C_{2}\m_{\kappa,\pm\gamma}(z_{\pm}),
\end{equation}
where $C_{1}$ and $C_{2}$ are arbitrary constant coefficients. In view of Eqs.~(\ref{varphi}),~(\ref{fpm}), and~(\ref{general}) the corresponding
solutions of Eq.~(\ref{phieq}) for positive and negative frequency solutions are
\begin{align}
\label{u}
U_{\k}(x)&=\Omega^{\frac{2-D}{2}}(\tau)e^{+i\k\cdot\x}\Big(C_{1}\wwp(z_{+})+C_{2}\wmp(z_{+})\Big), \\
\label{v}
V_{\k}(x)&=\Omega^{\frac{2-D}{2}}(\tau)e^{-i\k\cdot\x}\Big(C_{1}\wwm(z_{-})+C_{2}\wmm(z_{-})\Big),
\end{align}
where the choice of the sign for the $\gamma$ parameter will follow without ambiguity from the discussion in Sec.~\ref{sec:mode}.
%%%%%%%%%%%%%%%%%%%%%%%%%%%%%%%%%%%%%%%%%%%%%%%%%%%%%%%%%%%%%%%%%%%%%%%%%%%%%%%%%%%%%%%%%%%%%%%%%%%%%%%%%%%%%%%%%%%%%%%%%%%%%%%%%%%%%%%%%%%%%%%%%%%%%%%%%%%%%%%
%%%%%%%%%%%%%%%%%%%%%%%%%%%%%%%%%%%%%%%%%%%%%%%%%%%%%%%%%%%%%%%%%%%%%%%%%%%%%%%%%%%%%%%%%%%%%%%%%%%%%%%%%%%%%%%%%%%%%%%%%%%%%%%%%%%%%%%%%%%%%%%%%%%%%%%%%%%%%%%
\subsection{\label{sec:mode}Mode functions}
We need mode functions that determine the creation and annihilation operators and hence the vacuum state of the quantum field theory.
This vacuum will be determined by specifying the asymptotic form of the mode functions \cite{book:Birrell,book:Parker}.
In order to determine the mode functions at early times, which is approached as $t\rightarrow-\infty$, we impose that the functions
$f^{\pm}(z_{\pm})$, given by Eq.~(\ref{general}), asymptotically take the form $f^{\pm}(z_{\pm})\sim e^{\mp ik\tau}$ as $\tau\rightarrow-\infty$.
A comparison with the Minkowski spacetime mode functions shows that the functions $f^{+}(z_{+})$ and $f^{-}(z_{-})$ are positive and negative
frequency mode functions, respectively.
By the virtue of asymptotically expansions of the Whittaker functions $\wwp(z)$ and $\wmp(z)$ as $|z|\rightarrow\infty$ [see Eqs.~(\ref{win})
and~(\ref{min}), respectively], the normalized positive and negative frequency mode functions are \cite{Kim:2016dmm}, respectively,
\begin{align}
\label{uin}
U_{in\k}(x)&=(2k)^{-\frac{1}{2}}e^{\frac{i\pi\kappa}{2}}\Omega^{\frac{2-D}{2}}(\tau)e^{+i\k\cdot\x}\wwp(z_{+}), \\
\label{vin}
V_{in\k}(x)&=(2k)^{-\frac{1}{2}}e^{-\frac{i\pi\kappa}{2}}\Omega^{\frac{2-D}{2}}(\tau)e^{-i\k\cdot\x}\wwm(z_{-}).
\end{align}
A similar discussion is possible in the asymptotic future ($t \rightarrow \infty$). The desired asymptotic form is
$f^{\pm}(z_{\pm})\sim e^{\mp i|\gamma|Ht}$, leading with Eqs.~(\ref{mout}) and~(\ref{wout}) to the mode functions \cite{Kim:2016dmm}
\begin{align}
\label{uout}
U_{out\k}(x)&=(4|\gamma|k)^{-\frac{1}{2}}e^{\frac{i\pi\gamma}{2}}\Omega^{\frac{2-D}{2}}(\tau)e^{+i\k\cdot\x}\wmp(z_{+}),
\\ \label{vout}
V_{out\k}(x)&=(4|\gamma|k)^{-\frac{1}{2}}e^{\frac{i\pi\gamma}{2}}\Omega^{\frac{2-D}{2}}(\tau)e^{-i\k\cdot\x}\wmm(z_{-}).
\end{align}
The subscripts $in/out$ denote that these mode functions have the desired asymptotic form at early/late times, and the corresponding vacuum state
is referred to as the in vacuum and out vacuum, respectively.
\par
Since the orthonormality of the mode functions should be independent of time, there exists a conserved scalar product.
Between two scalar functions $u_{1}(x)$ and $u_{2}(x)$ it is defined in $D=1+d$ dimension by
\begin{equation}\label{produc}
\big(u_{1},u_{2}\big)=i\int d^{d}x\sqrt{|g|}g^{0\nu}\Big(u_{1}^{\ast}\rnd_{\nu}u_{2}-u_{2}\rnd_{\nu}u_{1}^{\ast}\Big),
\end{equation}
where the integral is taken over a constant $x^{0}$ hypersurface \cite{book:Birrell,book:Parker}.
If $u_{1}(x)$ and $u_{2}(x)$ are solutions of the field equation~(\ref{kgeq}) which vanish at spacial infinity, then $(u_{1},u_{2})$ is conserved \cite{book:Parker}. The mode functions~(\ref{uin})-(\ref{vout}) will be orthonormal with respect to the scalar product~(\ref{produc}) integrated
over a constant $\tau$ hypersurface. Then, subsequent orthonormality relations are satisfied
\begin{align}\label{orthonorm}
\big(U_{in(out)\k},U_{in(out)\k'}\big)&=-\big(V_{in(out)\k},V_{in(out)\k'}\big)
=(2\pi)^{d}\delta^{(d)}(\k-\k'), \nn\\
\big(U_{in(out)\k},V_{in(out)\k'}\big)&=0.
\end{align}
%%%%%%%%%%%%%%%%%%%%%%%%%%%%%%%%%%%%%%%%%%%%%%%%%%%%%%%%%%%%%%%%%%%%%%%%%%%%%%%%%%%%%%%%%%%%%%%%%%%%%%%%%%%%%%%%%%%%%%%%%%%%%%%%%%%%%%%%%%%%%%%%%%%%%%%%%%%%%%%
%%%%%%%%%%%%%%%%%%%%%%%%%%%%%%%%%%%%%%%%%%%%%%%%%%%%%%%%%%%%%%%%%%%%%%%%%%%%%%%%%%%%%%%%%%%%%%%%%%%%%%%%%%%%%%%%%%%%%%%%%%%%%%%%%%%%%%%%%%%%%%%%%%%%%%%%%%%%%%%
\section{\label{sec:creat}Particle creation}
In Sec.~\ref{sec:dS}, two complete sets of orthonormal mode functions were obtained, i.e., $\{U_{in\k},V_{in\k}\}$ given by
Eqs.~(\ref{uin}) and (\ref{vin}) and $\{U_{out\k},V_{out\k}\}$ given by Eqs.~(\ref{uout}) and (\ref{vout}).
In this section, we will derive a first result describing the Schwinger pair creation rate: the decay rate that we will derive via a Bogoliubov
transformation; see also \cite{Kim:2016dmm}.
Analogous methods were used to compute the pair creation rate in time-dependent field in Minkowski spacetime \cite{Gavrilov:1996pz} and without
an electric field for bosons in dS in \cite{Anderson:2013ila}.
In \cite{Kluger:1998bm} the connection of this Bogoliubov technique to kinetic theory was shown.
We will then compute the semiclassical conduction current.
The conduction current will be computed in the full generality in Sec.~\ref{sec:curent}, and a comparison with the semiclassical expression
will be performed.
\par
The scalar field operator $\phi(x)$ may be expanded in terms of the $\{U_{in\k},V_{in\k}\}$ set in the form
\begin{equation}\label{phiin}
\phi(x)=\int\frac{d^{d}k}{(2\pi)^{d}}\Big[U_{in\k}(x)a_{in\k}+V_{in\k}(x)b^{\dag}_{in\k}\Big],
\end{equation}
where $a_{in\k}$ annihilates particles described by the mode function $U_{in\k}$, and $b^{\dag}_{in\k}$ creates antiparticles described by the mode
function $V_{in\k}$. The quantization of the theory is implemented by adopting the commutation relations
\begin{equation}\label{comin}
\big[a_{in\k},a_{in\k'}^{\dag}\big]=\big[b_{in\k},b_{in\k'}^{\dag}\big]=(2\pi)^{d}\delta^{(d)}(\k-\k').
\end{equation}
The vacuum state is defined as
\begin{align}\label{vacin}
a_{in\k}|0\rangle_{in}=0, && \forall\,\k,
\end{align}
and then the construction of the Fock space can be done similarly to the Minkowski spacetime case. However, there is no $\rnd/\rnd x^{0}$ Killing
vector to define positive frequency mode functions, and consequently a unique mode decomposition of the scalar field operator $\phi(x)$ does not
exist. Therefore, $\phi(x)$ may be expanded in terms of a second complete set of orthonormal mode functions in the form
\begin{equation}\label{phiout}
\phi(x)=\int\frac{d^{d}k}{(2\pi)^{d}}\Big[U_{out\k}(x)a_{out\k}+V_{out\k}(x)b^{\dag}_{out\k}\Big],
\end{equation}
where $a_{out\k}$ annihilates particles described by the mode function $U_{out\k}$, and $b^{\dag}_{out\k}$ creates antiparticles described by the
mode function $V_{out\k}$. In this case, the commutation relations are
\begin{equation}\label{comout}
\big[a_{out\k},a_{out\k'}^{\dag}\big]=\big[b_{out\k},b_{out\k'}^{\dag}\big]=(2\pi)^{d}\delta^{(d)}(\k-\k').
\end{equation}
The decomposition of $\phi(x)$ in Eq.~(\ref{phiout}) defines a new vacuum state
\begin{align}\label{vacout}
a_{out\k}|0\rangle_{out}=0, && \forall\,\k,
\end{align}
and a new Fock space. Since both sets are complete, the orthonormal mode functions $U_{out\k}$ can be expanded in terms of the first complete
set of orthonormal mode functions. Hence
\begin{equation}\label{bogolu}
U_{out\k}(x)=\int \frac{d^{d}k'}{(2\pi)^{d}}\Big[\alpha_{\k,\k'}U_{in\k'}(x)+\beta_{\k,\k'}V_{in\k'}(x)\Big].
\end{equation}
By virtue of the orthonormality relations~(\ref{orthonorm}) the Bogoliubov coefficients $\alpha_{\k,\k'}$ and $\beta_{\k,\k'}$ will be determined by
\begin{align}\label{bogolab}
\alpha_{\k,\k'}&=\big(U_{out\k},U_{in\k'}\big), & \beta_{\k,\k'}&=-\big(U_{out\k},V_{in\k'}\big),
\end{align}
where the Bogoliubov coefficients satisfy the relations
\begin{align}\label{bogol}
\int\frac{d^{d}k}{(2\pi)^{d}}\Big[\alpha^{\ast}_{\k,\k'}\alpha_{\k,\k''}-\beta_{\k,\k'}\beta^{\ast}_{\k,\k''}\Big]
&=(2\pi)^{d}\delta^{(d)}\big(\k'-\k''\big), \nn\\
\int\frac{d^{d}k}{(2\pi)^{d}}\Big[\alpha^{\ast}_{\k,\k'}\beta_{\k,\k''}-\beta_{\k,\k'}\alpha^{\ast}_{\k,\k''}\Big]&=0.
\end{align}
As a consequence of Eqs.~(\ref{phiin}), (\ref{phiout}), and (\ref{bogolu}) the late time annihilation operator $a_{out\k}$ is related to the early
time annihilation operator $a_{in\k}$ by a Bogoliubov transformation
\begin{equation}\label{boga}
a_{out\k}=\int\frac{d^{d}k'}{(2\pi)^{d}}\Big[\alpha^{\ast}_{\k,\k'}a_{in\k'}-\beta^{\ast}_{\k,\k'}b^{\dag}_{in\k'}\Big].
\end{equation}
Using $a_{out\k}$ and the vacuum state $|0\rangle_{in}$ we can calculate the expectation value of the particle number operator\footnote{One can
verify that $_{in}\langle 0|a^{\dag}_{out\k}a_{out\k}|0\rangle_{in}=\,_{out}\langle 0|a^{\dag}_{in\k}a_{in\k}|0\rangle_{out}$.}
\begin{equation}\label{nout}
_{in}\langle 0|N_{out\k}|0\rangle_{in}=\,_{in}\langle 0|a^{\dag}_{out\k}a_{out\k}|0\rangle_{in}
=\int\frac{d^{d}k'}{(2\pi)^{d}}\big|\beta_{\k,\k'}\big|^{2}.
\end{equation}
Therefore, if $\big|\beta_{\k,\k'}\big|^{2}\neq0$, then particles are created.
%%%%%%%%%%%%%%%%%%%%%%%%%%%%%%%%%%%%%%%%%%%%%%%%%%%%%%%%%%%%%%%%%%%%%%%%%%%%%%%%%%%%%%%%%%%%%%%%%%%%%%%%%%%%%%%%%%%%%%%%%%%%%%%%%%%%%%%%%%%%%%%%%%%%%%%%%%%%%%%
%%%%%%%%%%%%%%%%%%%%%%%%%%%%%%%%%%%%%%%%%%%%%%%%%%%%%%%%%%%%%%%%%%%%%%%%%%%%%%%%%%%%%%%%%%%%%%%%%%%%%%%%%%%%%%%%%%%%%%%%%%%%%%%%%%%%%%%%%%%%%%%%%%%%%%%%%%%%%%%
\subsection{\label{sec:den}Density of created pairs}
In order to obtain the density of the created pairs an explicit expression of the Bogoliubov coefficients is needed.
To identify the Bogoliubov coefficients, the orthonormal mode functions, given by Eqs.~(\ref{uin})-(\ref{vout}), should be substituted into
Eq.~(\ref{bogolab}). We then obtain
\begin{align}
\label{alpha}
\alpha_{\k,\k'}&=(2\pi)^{d}\delta^{(d)}\big(\k-\k'\big)\alpha_{\k}, &
\alpha_{\k}&=(2|\gamma|)^{\frac{1}{2}}\frac{\Gamma(-2\gamma)}{\Gamma(\frac{1}{2}-\gamma-\kappa)}
e^{\frac{i\pi}{2}(\kappa-\gamma)}, \\
\label{beta}
\beta_{\k,\k'}&=(2\pi)^{d}\delta^{(d)}\big(\k+\k'\big)\beta_{\k}, &
\beta_{\k}&=-i(2|\gamma|)^{\frac{1}{2}}\frac{\Gamma(-2\gamma)}{\Gamma(\frac{1}{2}-\gamma+\kappa)}
e^{\frac{i\pi}{2}(\kappa+\gamma)},
\end{align}
and the normalization condition $|\alpha_{\k}|^{2}-|\beta_{\k}|^{2}=1$ is satisfied.
The expected number of the created pairs, with a given comoving momentum $\k$, in the in vacuum is given by Eq.~(\ref{nout}).
After a short calculation, Eq.~(\ref{beta}) results in
\begin{align}\label{betasq}
|\beta_{\k,\k'}|^{2}=\Big((2\pi)^{d}\delta^{(d)}(\k+\k')\Big)^{2}|\beta_{\k}|^{2}, &&
|\beta_{\k}|^{2}=\frac{e^{-2\pi|\gamma|}+e^{2\pi i\kappa}}{2 \sinh(2\pi|\gamma|)}.
\end{align}
For convenience we normalize the $d$ volume of $\dsd$ in a box with dimensions $L^{d}$.
Then, the number of created pairs per comoving $d$ volume, with given comoving momentum $\k$ is
\begin{equation}\label{nv}
\frac{1}{L^{d}}\times\int\frac{d^{d}k'}{(2\pi)^{d}}\big|\beta_{\k,\k'}\big|^{2}=|\beta_{\k}|^{2}.
\end{equation}
Using the mathematical formulas~(\ref{formul}), (\ref{area}), and (\ref{areaint}), the number of created pairs per unit $d$ volume is
\begin{equation}\label{intk}
\int\frac{d^{d}k}{(2\pi)^{d}}|\beta_{\k}|^{2}
=\frac{1}{(4\pi)^{\frac{d}{2}}\sinh(2\pi|\gamma|)}\Big(\frac{e^{-2\pi|\gamma|}}{\Gamma(\frac{d}{2})}
+(\pi\lambda)^{1-\frac{d}{2}}\I_{\frac{d}{2}-1}(2\pi\lambda)\Big)
\int_{0}^{\infty}k^{d-1}\,dk,
\end{equation}
where $\I_{\nu}$ is the modified Bessel function; see Appendix~\ref{app:bessel}.
This integral is not finite, since it takes into account the total number of created pairs from the infinite past to the infinite future.
However, the number of created pairs per unit of time is finite. Thus, we convert the $k$ integral in Eq.~(\ref{intk}) into a $\tau$ integral.
\par
To have physically acceptable particles states, one needs to have particle states well defined in the asymptotic past and future; that is, the
background gravitational and electric fields vary slowly.
This is called the adiabatic condition and is a semiclassical approximation.
In the case of positive frequencies, the mode equation~(\ref{feq}) can be rewritten
\begin{equation}\label{refeq}
\frac{d^{2}f(\tau)}{d\tau^{2}}+\omega^{2}f(\tau)=0,
\end{equation}
where the momentum dependent frequency is
\begin{equation}\label{omega}
\omega^{2}=k^{2}-\frac{2eEk_{x}}{H^{2}\tau}+\frac{1}{\tau^{2}}\Big(\frac{m^{2}}{H^{2}}+\frac{e^{2}E^{2}}{H^{4}}
+\frac{1-d^{2}}{4}+\xi D(D-1)\Big),
\end{equation}
and we have used $R=D(D-1)H^{2}$. The adiabatic condition requires that at all times, the frequency $\omega$ satisfies the relations
\begin{align}\label{condition}
\frac{\dot{\omega}^{2}}{\omega^{4}}&\ll 1, & \frac{\ddot{\omega}}{\omega^{3}}&\ll 1,
\end{align}
where dots refer to the partial derivative with respect to the conformal time $\tau$.
In the infinite past $\tau\rightarrow-\infty$ the frequency approaches $\omega\rightarrow k$, and hence the adiabatic condition~(\ref{condition})
is satisfied. In the infinite future $\tau\rightarrow 0$ we have
\begin{equation}\label{omegao}
\frac{\dot{\omega}^{2}}{\omega^{4}}\sim\frac{1}{2}\frac{\ddot{\omega}}{\omega^{3}}\sim
\Big(\lambdam^2+\lambda^2 +\frac{1-d^{2}}{4}\Big)^{-1},
\end{equation}
under the condition that
\begin{equation}\label{assumption}
\lambdam^{2}+\lambda^{2}\gg \frac{d^2-1}{4},
\end{equation}
the adiabatic condition is satisfied.
In our investigation, the spacetime dimension is not too large, i.e., $d\sim 1$. Hence, the condition~(\ref{assumption}) implies
\begin{equation}\label{assume}
\rho^{2}=\lambdam^{2}+\lambda^{2}\simeq|\gamma|^{2} \gg 1.
\end{equation}
The condition~(\ref{assumption}) justifies our assumption about the range of parameters $\lambdam$ and $\lambda$ which leads to $\gamma^{2}<0$.
Observe that assuming~(\ref{assumption}) implies that, if one does not want to have trivial flat spacetime results, one needs to assume as well
\begin{equation}\label{lambdamgg}
\lambdam^{2} \gg \frac{d^{2}-1}{4}.
\end{equation}
The proof of the previous statement is that if one assumes $\lambdam\ll 1$, together with the semiclassical condition~(\ref{assume}), it will be
equivalent to assume $\lambda\gg 1$.
In this limit the scalar field, the electromagnetic field, and the de~Sitter spacetime are conformally invariant, leading to flat spacetime results;
see the discussion in Sec.~\ref{sec:ur}.
Two regimes can be discussed then under the semiclassical approximation: strong electric field, $\lambda\gg\max(1,\lambdam)$ which will give the
flat spacetime results, and heavy scalar field, $\lambdam\gg\max(1,\lambda)$.
\par
Under~(\ref{assume}), i.e., $\rho\gg 1$, it can be verified \cite{Frob:2014zka} that the extremum of $|\dot{\omega}/\omega^{2}|$ occurs around
the time
\begin{equation}\label{extremum}
\tau\sim-\frac{\rho}{k}.
\end{equation}
More on converting momentum to the time integral, in the context of Schwinger pair creation, can be found in \cite{Gavrilov:1996pz,Anderson:2013ila,Kluger:1998bm}.
As a consequence of Eq.~(\ref{extremum}), the $k$ integral in the Eq.~(\ref{intk}) can be converted into a $\tau$ integral
\begin{eqnarray}\label{intt}
\int\frac{d^{d}k}{(2\pi)^{d}}|\beta_{\k}|^{2}&=&
\frac{1}{(4\pi)^{\frac{d}{2}}\sinh(2\pi\rho)}\Big(\frac{e^{-2\pi\rho}}{\Gamma(\frac{d}{2})}
+(\pi\lambda)^{1-\frac{d}{2}}\I_{\frac{d}{2}-1}(2\pi\lambda)\Big) \nn\\
&\times&H^{D}\rho^{d}\int_{-\infty}^{0}\Omega^{D}(\tau)d\tau.
\end{eqnarray}
The number of created pairs per unit of $D$ volume of $\dsd$ or the decay rate is then given by
\begin{eqnarray}\label{rate}
\Gamma&:=&\frac{1}{\Delta V}\times\int\frac{d^{d}k}{(2\pi)^{d}}|\beta_{\k}|^{2} \nn\\
&=&\frac{H^{D}\rho^{d}}{(4\pi)^{\frac{d}{2}}\sinh(2\pi\rho)}
\Big(\frac{e^{-2\pi\rho}}{\Gamma(\frac{d}{2})}+(\pi\lambda)^{1-\frac{d}{2}}
\I_{\frac{d}{2}-1}(2\pi\lambda)\Big),
\end{eqnarray}
where
\begin{equation}\label{slice}
\Delta V=\Omega^{D}(\tau)\Delta\tau,
\end{equation}
is the slice of $D$ volume in the conformal time interval $\Delta\tau$.
Using Eqs.~(\ref{cosh}) and~(\ref{sinh}), it can be shown that in the cases $D=2$ and $D=4$, Eq.~(\ref{rate}) gives the same result as
\cite{Frob:2014zka} and \cite{Kobayashi:2014zza}, respectively.
The decay rate~(\ref{rate}) is independent of time, and as a consequence the number density in the comoving frame at time $\tau$ reads
\begin{equation}\label{density}
n=\Omega^{-d}(\tau)\int_{-\infty}^{\tau}d\tau'\Omega^{D}(\tau')\Gamma=\frac{\Gamma}{Hd}.
\end{equation}
It is constant with respect to time; therefore, the number of pairs created by the background electric and gravitational fields is exactly balanced
by the expansion of the dS.
\par
Provided that Eq.~(\ref{assume}) is satisfied, then the Bogoliubov coefficient~(\ref{betasq}) is approximated as
\begin{equation}\label{betaapprox}
|\beta_{\k}|^{2}\simeq e^{-4\pi\rho}+e^{-2\pi(\rho-\lambda r)}.
\end{equation}
From the definitions in Eq.~(\ref{lambda}), $|r|\leq 1$ implying $\rho\geq\lambda r$, so the first term in the right-hand side of
Eq.~(\ref{betaapprox}) is smaller than the second one. Then, to leading order, we find
\begin{equation}\label{approxbeta}
|\beta_{\k}|^{2}\simeq\exp\Big[-2\pi\Big(\Big(\frac{\mds^{2}}{H^{2}}+
\frac{(eE)^{2}}{H^{4}}\Big)^{\frac{1}{2}}+\frac{eE}{H^{2}}\frac{k_{x}}{k}\Big)\Big].
\end{equation}
Therefore, under the semiclassical condition~(\ref{assume}), $\beta_{\k}$ is nonzero for both $k_{x}>0$ and $k_{x}<0$ \cite{Garriga:1994bm}.
In the language of nucleation of bubbles, considering $E>0$ and taking the particle with charge $|e|$ to the right of the particle with charge
$-|e|$, the pairs can nucleate in both the screening and the antiscreening orientations (corresponding to $k_{x}<0$ and $k_{x}>0$, respectively)
because of the gravitational effects \cite{Garriga:1993fh}.
Hence, creating charges in the screening orientation tends to decrease the background electrical field while creating them in the antiscreening
orientation tends to increase it.
Usually screening and antiscreening orientations are referred to as downward and upward tunneling \cite{Frob:2014zka}.
\par
The Minkowski spacetime limit is obtained in the limit $H\rightarrow0$. The decay rate~(\ref{rate}) in this limit approaches
\begin{equation}\label{limit}
\lim_{H\rightarrow0}\Gamma=\frac{|eE|^{\frac{D}{2}}}{(2\pi)^{d}}e^{-\frac{\pi m^{2}}{|eE|}},
\end{equation}
which is the same result with the Schwinger pair production rate in $D$-dimensional Minkowski spacetime \cite{Gavrilov:1996pz}.
In $\dsd$, the pair production rate is higher than in flat spacetime, due to the gravitational pair production contribution.
%%%%%%%%%%%%%%%%%%%%%%%%%%%%%%%%%%%%%%%%%%%%%%%%%%%%%%%%%%%%%%%%%%%%%%%%%%%%%%%%%%%%%%%%%%%%%%%%%%%%%%%%%%%%%%%%%%%%%%%%%%%%%%%%%%%%%%%%%%%%%%%%%%%%%%%%%%%%%%%
%%%%%%%%%%%%%%%%%%%%%%%%%%%%%%%%%%%%%%%%%%%%%%%%%%%%%%%%%%%%%%%%%%%%%%%%%%%%%%%%%%%%%%%%%%%%%%%%%%%%%%%%%%%%%%%%%%%%%%%%%%%%%%%%%%%%%%%%%%%%%%%%%%%%%%%%%%%%%%%
\subsection{\label{sec:semc}Semiclassical current}
Because of the electrical field, the newly created pairs start to move and hence create a conductive current.
In this subsection we present a first semiclassical expression for it, following similar steps as in \cite{Frob:2014zka,Kobayashi:2014zza}.
In general, the relation between the current $J_{\sem}$ and the density $n$ of the semiclassical particles with charge $e$ and velocity $v$ is
$J_{\sem}=2evn$. The density of created pairs can be read from Eqs.~(\ref{rate}) and~(\ref{density}).
Hence, the semiclassical current is determined by
\begin{eqnarray}\label{semi}
J_{\sem}&=&\frac{2e\Gamma}{Hd} v \nn\\
&=&\frac{2evH^{d}\rho^{d}}{(4\pi)^{\frac{d}{2}}d\sinh\big(2\pi\rho\big)}
\bigg(\frac{e^{-2\pi\rho}}{\Gamma(\frac{d}{2})}+\big(\pi\lambda\big)^{1-\frac{d}{2}}
\I_{\frac{d}{2}-1}\big(2\pi\lambda\big)\bigg).
\end{eqnarray}
For a semiclassical particle with a comoving momentum $k_{i}$ that interacts with the background vector potential~(\ref{vector}), the components
of the physical momentum vector $p^{\mu}_{\k}$ can be written as
\begin{eqnarray}\label{momentum}
p^{0}_{\k}&=&\Omega^{-1}(\tau)\Big(\mds^{2}+\Omega^{-2}(\tau)\delta^{ij}(k_{i}+eA_{i})(k_{j}+eA_{j})\Big)^{\frac{1}{2}}, \nn\\
p^{i}_{\k}&=&\Omega^{-2}(\tau)\delta^{ij}(k_{j}+eA_{j}), \hspace{1cm} i=1,\ldots,d,
\end{eqnarray}
and then the magnitude of the velocity reads $v=|\mathbf{p}_{\k}|/p^{0}_{\k}$.
%%%%%%%%%%%%%%%%%%%%%%%%%%%%%%%%%%%%%%%%%%%%%%%%%%%%%%%%%%%%%%%%%%%%%%%%%%%%%%%%%%%%%%%%%%%%%%%%%%%%%%%%%%%%%%%%%%%%%%%%%%%%%%%%%%%%%%%%%%%%%%%%%%%%%%%%%%%%%%%
%%%%%%%%%%%%%%%%%%%%%%%%%%%%%%%%%%%%%%%%%%%%%%%%%%%%%%%%%%%%%%%%%%%%%%%%%%%%%%%%%%%%%%%%%%%%%%%%%%%%%%%%%%%%%%%%%%%%%%%%%%%%%%%%%%%%%%%%%%%%%%%%%%%%%%%%%%%%%%%
\subsubsection{\label{sec:ur}Strong electric field regime}
In the strong electric field regime, the relation $\lambda\gg\max(1,\lambdam)$ is satisfied.
Using Eqs.~(\ref{extremum}) and~(\ref{momentum}) one can show that
\begin{eqnarray}\label{momentums}
p^{0}_{\k}&=&\Omega^{-1}(\tau)H\lambda\sqrt{2(1+\cos\theta_{1})}, \nn\\
p^{1}_{\k}&=&-\Omega^{-1}(\tau)H\lambda(1+\cos\theta_{1}), \nn\\
p^{i}_{\k}&=&-\Omega^{-1}(\tau)H\lambda\omega^{i}, \hspace{1cm} i=2,\ldots,d,
\end{eqnarray}
where $\omega^{i}$ is given by Eq.~(\ref{cordinat}), and consequently $v\sim 1$.
Hence, in the strong electric field regime the created particles are ultrarelativistic.
Using an asymptotic expansion of the modified Bessel function~(\ref{infty}), it can be shown that in the strong electric field regime,
$\lambda\gg\max(1,\lambdam)$, Eq.~(\ref{semi}) is approximated as
\begin{equation}\label{semiur}
J_{\sem}\simeq\frac{2e|eE|^{\frac{D}{2}}}{H(2\pi)^{d}d}e^{-\frac{\pi\mds^{2}}{|eE|}}.
\end{equation}
An explicit comparison to the flat spacetime is possible in the strong electric field regime.
Indeed, under the same assumptions, the flat spacetime limit in $D$ dimensions reads
\begin{equation}\label{semij}
J_{\mathrm{sem,flat}}\simeq\frac{2et|eE|^{\frac{D}{2}}}{(2\pi)^{d}d}e^{-\frac{\pi m^{2}}{|eE|}},
\end{equation}
with $t$ being the Minkowski time \cite{Anderson:2013ila,Anderson:2013zia}.
In the expanding dS, accounting for the spacetime dilution in the comoving frame can be done by substituting $t\rightarrow H^{-1}$.
In the limit $|\lambda|\rightarrow\infty$, the exponential factor in Eq.~(\ref{semiur}) approaches unity; then $J_{\sem}$ becomes independent
of the scalar field mass and responds as $E^{\frac{D}{2}}$.
The result corresponds to the result of massless scalar fields in flat spacetime.
%%%%%%%%%%%%%%%%%%%%%%%%%%%%%%%%%%%%%%%%%%%%%%%%%%%%%%%%%%%%%%%%%%%%%%%%%%%%%%%%%%%%%%%%%%%%%%%%%%%%%%%%%%%%%%%%%%%%%%%%%%%%%%%%%%%%%%%%%%%%%%%%%%%%%%%%%%%%%%%
%%%%%%%%%%%%%%%%%%%%%%%%%%%%%%%%%%%%%%%%%%%%%%%%%%%%%%%%%%%%%%%%%%%%%%%%%%%%%%%%%%%%%%%%%%%%%%%%%%%%%%%%%%%%%%%%%%%%%%%%%%%%%%%%%%%%%%%%%%%%%%%%%%%%%%%%%%%%%%%
\subsubsection{\label{sec:nr}Heavy scalar field regime}
In the heavy scalar field regime, the relation $\lambdam\gg\max(1,\lambda)$ is satisfied.
It was shown in \cite{Frob:2014zka} that due to the background electric field, the charged particles have a terminal physical momentum at late
times which is determined as
\begin{equation}\label{terminalm}
p^{i}_{\k}=\Omega^{-1}(\tau)\lim_{\tau\rightarrow 0}\Big(-H\tau k_{i}+\frac{eE}{H}\delta_{i,1}\Big)
=-\Omega^{-1}(\tau)H\lambda\delta_{i,1}, \hspace{1cm} i=1,\ldots,d.
\end{equation}
In the heavy scalar field regime, we consider the terminal value for the physical momentum that leads to
\begin{equation}\label{terminale}
p^{0}_{\k}\simeq\Omega^{-1}(\tau)H\lambdam,
\end{equation}
and consequently the terminal velocity becomes $v\sim|\lambda|/\lambdam$ \cite{Frob:2014zka}.
Hence, in the heavy scalar field regime, $\lambdam\gg\max(1,\lambda)$, the leading order term of the expansion of Eq.~(\ref{semi}) is obtained as
\begin{eqnarray}\label{seminr}
J_{\sem}\simeq\frac{2eH^{d-3}\mds^{3}}{(2\pi)^{d-1}d}\Big|\frac{eE}{\mds^{2}}\Big|^{\frac{4-d}{2}}
\I_{\frac{d}{2}-1}(2\pi\lambda)e^{-2\pi\lambdam}.
\end{eqnarray}
Thus, for heavy, i.e., nonrelativistic charged particles the semiclassical current is exponentially suppressed.
\par
We will give a more rigorous derivation of these results in $D=3$ dimension below: we will see that, in the strong electric field regime,
the semiclassical current $J_{\sem}$ agrees with the expectation value of the current operator, whereas in the heavy scalar field regime, they
are exponentially different from each other.
%%%%%%%%%%%%%%%%%%%%%%%%%%%%%%%%%%%%%%%%%%%%%%%%%%%%%%%%%%%%%%%%%%%%%%%%%%%%%%%%%%%%%%%%%%%%%%%%%%%%%%%%%%%%%%%%%%%%%%%%%%%%%%%%%%%%%%%%%%%%%%%%%%%%%%%%%%%%%%%
%%%%%%%%%%%%%%%%%%%%%%%%%%%%%%%%%%%%%%%%%%%%%%%%%%%%%%%%%%%%%%%%%%%%%%%%%%%%%%%%%%%%%%%%%%%%%%%%%%%%%%%%%%%%%%%%%%%%%%%%%%%%%%%%%%%%%%%%%%%%%%%%%%%%%%%%%%%%%%%
\section{\label{sec:curent}Induced current and conductivity in $D=3$ dimension}
In this section, we confine ourselves to the case of $\ds_{3}$ and compute the induced current and the conductivity without imposing the condition
indicated in Eq.~(\ref{assume}).
Whereas the particle number has no meaning when the adiabatic future does not exist, the current is well defined and is indeed the right quantity
to describe the Schwinger effect in this context.
It can be shown that the current operator of the charged scalar field
\begin{equation}\label{operator}
j^{\mu}(x)=\frac{ie}{2}g^{\mu\nu}\Big(\{(\rnd_{\nu}\phi+ieA_{\nu}\phi),\phi^{\ast}\}-\{(\rnd_{\nu}\phi^{\ast}
-ieA_{\nu}\phi^{\ast}),\phi\}\Big),
\end{equation}
is conserved, i.e., $\nabla_{\mu}j^{\mu}=0$ \cite{book:Parker}.
Using Eqs.~(\ref{phiin})-(\ref{vacout}) it can be shown that in the in vacuum and out vacuum, $\langle j^{0}\rangle=0$.
However, in the in-vacuum state, the expectation of the spacelike component of the current operator is
\begin{align}\label{vev}
\langle j^{1}\rangle_{in}&=\,_{in}\langle0|j^{1}|0\rangle_{in} \nn\\
&=2e\Omega^{-3}(\tau)\int\frac{d^{2}k}{(2\pi)^{2}}\Big(k_{x}+eA_{1}(\tau)\Big)
\frac{e^{\kappa\pi i}}{2k}\big|\wwp(z_{+})\big|^{2}.
\end{align}
In order to compute the vacuum expectation value of the current operator~(\ref{operator}) we choose the in-vacuum state because this state is
Hadamard \cite{Garriga:1994bm,Frob:2014zka}.
Hence, the expectation value has a UV behavior similar to the flat spacetime.
Substituting explicit expressions, the integral~(\ref{vev}) can be rewritten as
\begin{align}
\langle j^{1}\rangle_{in}&=\frac{e}{2\pi^{2}}H^{2}\Omega^{-1}(\tau) \nn\\
&\times\lim_{\Lambda\rightarrow\infty}\int_{-1}^{1}\frac{dr}{\sqrt{1-r^{2}}}
\int_{0}^{\Lambda}dp\big(rp-\lambda\big)e^{\lambda r\pi}\big|\w_{-i\lambda r,\gamma}(-2ip)\big|^{2}, \label{curent}
\end{align}
where $\Lambda=-K\tau$ and $K$ is an upper cutoff on momentum $k$ introduced for convenience and that will be taken to infinity at the end of the
calculation. We also have introduced
\begin{equation}\label{pk}
p=-k\tau.
\end{equation}
The details of computation of the integral~(\ref{curent}) are reviewed in Appendix~\ref{app:int}. The final result is
\begin{align}\label{unreg}
&\langle j^{1}\rangle_{in}=\frac{e}{2\pi^{2}}H^{2}\Omega^{-1}(\tau)\bigg[-\frac{\pi}{2}\lambda
\lim_{\Lambda\rightarrow\infty}\Lambda \nn\\
&+\frac{\pi}{4}\lambda\gamma\cot(2\pi\gamma)+\frac{\gamma}{4\sin(2\pi\gamma)}\Big(3\I_{1}(2\pi\lambda)
-2\pi\lambda \I_{0}(2\pi\lambda)\Big)+\frac{i}{2\sin(2\pi\gamma)} \nn\\
&\times\int_{-1}^{1}\frac{dr}{\sqrt{1-r^{2}}}b_{r}\Big\{\big(e^{2\pi\lambda r}
+e^{-2\pi i\gamma}\big)\psi(\frac{1}{2}+i\lambda r-\gamma)-\big(e^{2\pi\lambda r}
+e^{2\pi i\gamma}\big)\psi(\frac{1}{2}+i\lambda r+\gamma)\Big\}\bigg],
\end{align}
where $\psi$ denotes the digamma function and the coefficient $b_{r}$ is defined as
\begin{equation}\label{br}
b_{r}=-\frac{3}{2}\lambda^{2}r^{3}+\Big(\frac{1}{8}-\frac{\gamma^{2}}{2}+\lambda^{2}\Big)r.
\end{equation}
%%%%%%%%%%%%%%%%%%%%%%%%%%%%%%%%%%%%%%%%%%%%%%%%%%%%%%%%%%%%%%%%%%%%%%%%%%%%%%%%%%%%%%%%%%%%%%%%%%%%%%%%%%%%%%%%%%%%%%%%%%%%%%%%%%%%%%%%%%%%%%%%%%%%%%%%%%%%%%%
%%%%%%%%%%%%%%%%%%%%%%%%%%%%%%%%%%%%%%%%%%%%%%%%%%%%%%%%%%%%%%%%%%%%%%%%%%%%%%%%%%%%%%%%%%%%%%%%%%%%%%%%%%%%%%%%%%%%%%%%%%%%%%%%%%%%%%%%%%%%%%%%%%%%%%%%%%%%%%%
\subsection{\label{sec:ren}Adiabatic subtraction}
In order to remove the UV divergence term from the expression~(\ref{unreg}) we need to apply a renormalization scheme.
In the context of quantum field theory in curved spacetime various regularization and renormalization methods have been developed \cite{book:Birrell,book:Parker}. The adiabatic subtraction or regularization method is achieved by subtracting terms computed in the limit of
slowly varying backgrounds to obtain a finite expression.
The idea of slow varying backgrounds is implemented by introducing adiabatic orders which in our problem will be nothing but counting time
derivatives in a given quantity.
More details about adiabatic subtraction in the context of Schwinger pair creation in curved spacetime are given in Ref.~\cite{Stahl:2015gaa}.
We will hence perform the adiabatic expansion of the mode functions up to the minimal order which makes the original expression~(\ref{unreg})
finite. To do so, we express the solution of the mode equation~(\ref{refeq}) as a Wentzel-Kramers-Brillouin (WKB) type solution
\begin{equation}\label{fa}
f_{\mathrm{A}}(\tau)=\big(2W(\tau)\big)^{-\frac{1}{2}}\exp\Big[-i\int^{\tau}W(\tau')d\tau'\Big],
\end{equation}
where in order to fulfill Eq.~(\ref{refeq}), the function $W$ satisfies the equation
\begin{equation}\label{weq}
W^{2}(\tau)=\omega^{2}(\tau)+\frac{3}{4}\frac{\dot{W}^{2}}{W^{2}}-\frac{1}{2}\frac{\ddot{W}}{W}.
\end{equation}
Provided that the adiabatic condition~(\ref{condition}) holds, derivative terms in Eq.~(\ref{weq}) will be negligible compared to $\omega^{2}$
terms. As we will see in this subsection, the zeroth order of the adiabatic expansion is enough to remove the UV divergent term from~(\ref{unreg}).
The zeroth order adiabatic expansion of $W$ is
\begin{equation}\label{wzero}
W^{(0)}(\tau)=\omega_{0}(\tau),
\end{equation}
where the superscript denotes the adiabatic order. The last term in $\omega^{2}$ [see Eq.~(\ref{omega})] can be rewritten in the form
\begin{equation}
\frac{6}{\tau^{2}}\Big(\xi-\frac{1}{8}\Big)=6\Big(\xi-\frac{1}{8}\Big)\frac{\dot{\Omega}^{2}}{\Omega^{2}},
\label{last}
\end{equation}
revealing that this term is of adiabatic order 2. Therefore, $\omega_{0}$ in Eq.~(\ref{wzero}) is given by
\begin{equation}\label{omegaz}
\omega_{0}(\tau)=+\Big(k^{2}-\frac{2eE}{H^{2}\tau}k_{x}+\frac{m^{2}}{H^{2}\tau^{2}}+
\frac{e^{2}E^{2}}{H^{4}\tau^{2}}\Big)^{\frac{1}{2}}.
\end{equation}
By virtue of Eqs.~(\ref{varphi}), (\ref{fpm}), (\ref{fa}), (\ref{wzero}), and (\ref{omegaz}), the zeroth order adiabatic expansion of the positive
frequency $U_{\mathrm{A}}$ and of the negative frequency $V_{\mathrm{A}}$ mode functions are
\begin{eqnarray}\label{uava}
U_{\mathrm{A};\k}(x)&=&\Omega^{-\frac{1}{2}}(\tau)\big(2\omega_{0}\big)^{-\frac{1}{2}}
\exp\Big[i\k\cdot\x-i\int^{\tau}\omega_{0}(\tau')d\tau'\Big], \nn\\
V_{\mathrm{A};-\k}(x)&=&\Omega^{-\frac{1}{2}}(\tau)\big(2\omega_{0}\big)^{-\frac{1}{2}}
\exp\Big[i\k\cdot\x+i\int^{\tau}\omega_{0}(\tau')d\tau'\Big].
\end{eqnarray}
We use this complete set of orthonormal mode functions to expand the charged scalar field operator, then substituting into Eq.~(\ref{operator})
leads to the zeroth order adiabatic expansion of the current operator
\begin{equation}\label{ja}
\langle j^{1}\rangle_{\mathrm{A}}=-\frac{eH^{2}}{4\pi}\lambda\Omega^{-1}(\tau)
\lim_{\Lambda\rightarrow\infty}\Lambda.
\end{equation}
We emphasize that in the expression~(\ref{ja}) there is no finite term or $\Lambda$-independent contribution.
Applying the adiabatic subtraction scheme,
\begin{eqnarray}\label{scheme}
\langle j^{1}\rangle_{\mathrm{reg}}&=&\langle j^{1}\rangle_{in}-\langle j^{1}\rangle_{\mathrm{A}} \nn\\
&=&\Omega^{-1}(\tau)J,
\end{eqnarray}
gives the regularized current as
\begin{align}\label{reg}
&J=\frac{eH^{2}}{8\pi^{2}}\frac{\gamma}{\sin(2\pi\gamma)}\bigg[\pi\lambda\cos(2\pi\gamma)
+3\I_{1}(2\pi\lambda)-2\pi\lambda\I_{0}(2\pi\lambda)+\frac{2i}{\gamma} \nn\\
&\times\int_{-1}^{1}\frac{b_{r}dr}{\sqrt{1-r^{2}}}\Big\{\big(e^{2\pi\lambda r}
+e^{-2\pi i\gamma}\big)\psi\big(\frac{1}{2}+i\lambda r-\gamma\big)
-\big(e^{2\pi\lambda r}+e^{2\pi i\gamma}\big)\psi\big(\frac{1}{2}+i\lambda r+\gamma\big)\Big\}\bigg].
\end{align}
By virtue of the modified Bessel function property given in Eq.~(\ref{parity}), one can show that $J$ is an odd function under the transformation $\lambda\rightarrow-\lambda$, illustrating that if one inverts the electrical field sense, the particles move in the opposite direction.
%%%%%%%%%%%%%%%%%%%%%%%%%%%%%%%%%%%%%%%%%%%%%%%%%%%%%%%%%%%%%%%%%%%%%%%%%%%%%%%%%%%%%%%%%%%%%%%%%%%%%%%%%%%%%%%%%%%%%%%%%%%%%%%%%%%%%%%%%%%%%%%%%%%%%%%%%%%%%%%
%%%%%%%%%%%%%%%%%%%%%%%%%%%%%%%%%%%%%%%%%%%%%%%%%%%%%%%%%%%%%%%%%%%%%%%%%%%%%%%%%%%%%%%%%%%%%%%%%%%%%%%%%%%%%%%%%%%%%%%%%%%%%%%%%%%%%%%%%%%%%%%%%%%%%%%%%%%%%%%
\subsection{\label{sec:result}Regularized current and conductivity}
\begin{figure}
\includegraphics[scale=0.9]{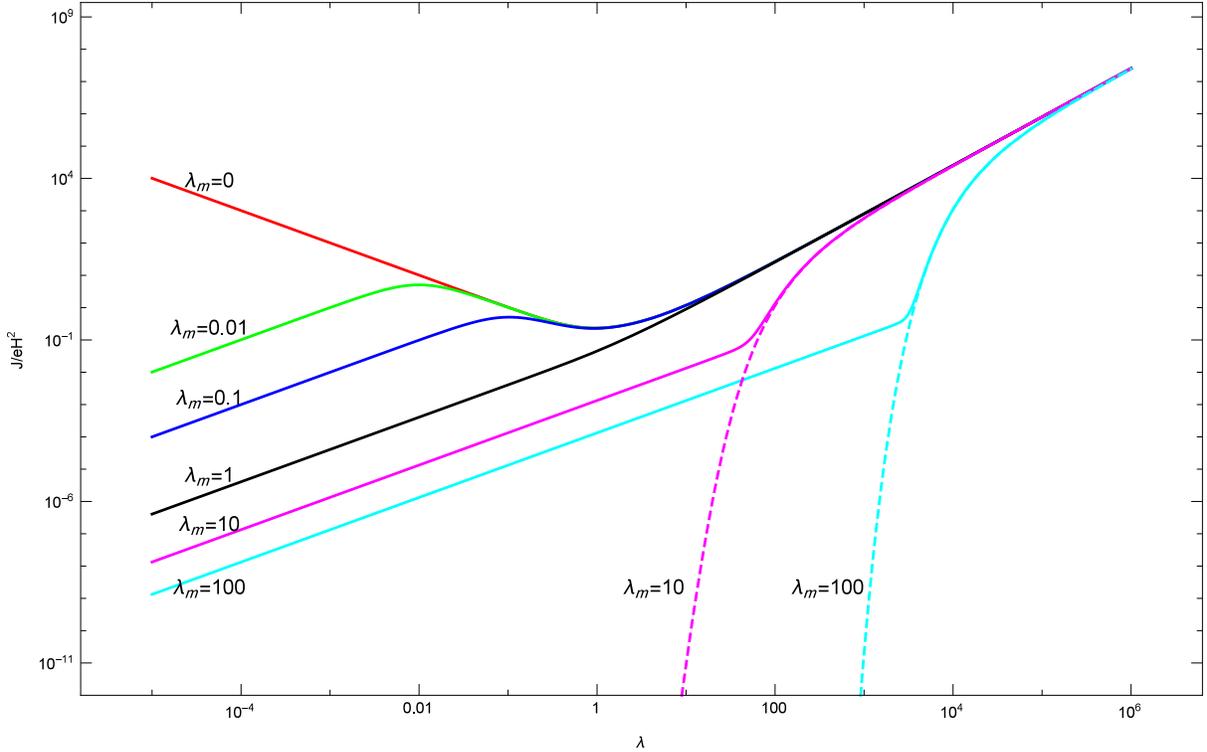}
\caption{For different values of $\lambdam$, the normalized quantum vacuum expectation value of the induced current $J/eH^{2}$ and the
semiclassical current $J_{\sem}/eH^{2}$ in $D=3$ dimension are plotted as a function of $\lambda$ with solid and dashed lines, respectively.}
\label{fig1}
\end{figure}

After computing the renormalized current, we consider the conductivity defined as
\begin{equation}\label{sigma}
\sigma:=\frac{J}{E}.
\end{equation}
We present a plot of the current~(\ref{reg}) and of the conductivity~(\ref{sigma}) in Figs.~\ref{fig1} and~\ref{fig2}, respectively.
The general features of these figures are that in the strong electric field regime $\lambda\gg\max(1,\lambdam)$ all the curves have
the same asymptotic behavior, and in the weak electric field regime $\lambda\ll\min(1,\lambdam)$ the current and conductivity are
suppressed for increasing scalar field mass.
For the case of a massless minimally coupled scalar field, i.e., $\lambdam=0$, for $\lambda\lesssim1$, the current and conductivity are increasing
as the electric field is decreasing.
This phenomenon was dubbed infrared-hyperconductivity (IR-HC) in \cite{Frob:2014zka}.
In the following subsections, we analytically investigate the limiting behaviors of the current and the conductivity.
In this analysis, for simplicity, we use the sign conventions $\lambda=|\lambda|$ and $J=|J|$.
%%%%%%%%%%%%%%%%%%%%%%%%%%%%%%%%%%%%%%%%%%%%%%%%%%%%%%%%%%%%%%%%%%%%%%%%%%%%%%%%%%%%%%%%%%%%%%%%%%%%%%%%%%%%%%%%%%%%%%%%%%%%%%%%%%%%%%%%%%%%%%%%%%%%%%%%%%%%%%%
%%%%%%%%%%%%%%%%%%%%%%%%%%%%%%%%%%%%%%%%%%%%%%%%%%%%%%%%%%%%%%%%%%%%%%%%%%%%%%%%%%%%%%%%%%%%%%%%%%%%%%%%%%%%%%%%%%%%%%%%%%%%%%%%%%%%%%%%%%%%%%%%%%%%%%%%%%%%%%%
\subsubsection{\label{sec:strong}Strong electric field regime}
\begin{figure}
\includegraphics[scale=0.9]{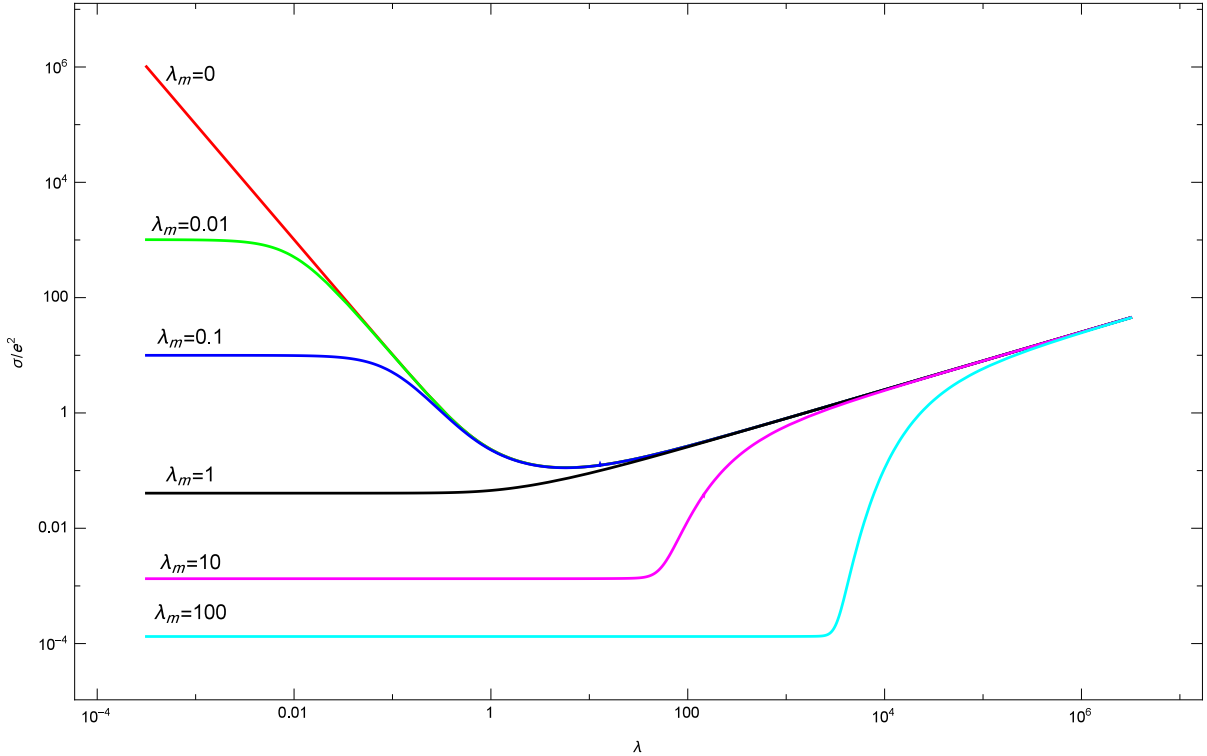}
\caption{For different values of $\lambdam$, the normalized conductivity $\sigma/e^{2}$ is plotted as a function of $\lambda$.
The phenomenon of infrared hyperconductivity appears for $\lambdam<\sqrt{3/4}$.} \label{fig2}
\end{figure}

Taking $\lambda\rightarrow\infty$ in the current expression~(\ref{reg}) with $\lambdam$ fixed, the leading order term is
\begin{align}\label{strong}
J&\simeq\frac{e^{2}}{4\pi^{2}}\frac{|eE|^{\frac{1}{2}}}{H}E, &
\sigma&\simeq\frac{e^{2}}{4\pi^{2}}\frac{|eE|^{\frac{1}{2}}}{H}.
\end{align}
The results~(\ref{strong}) analytically describe the behaviors of the current and conductivity shown by Figs.~\ref{fig1} and~\ref{fig2},
respectively. As illustrated in the figures, in this limit, the current and conductivity become increasing functions of electric field $E$ and
independent of $\mds$.
In the cases of $\ds_{2}$ \cite{Frob:2014zka} and $\ds_{4}$ \cite{Kobayashi:2014zza}, the authors showed that the current responds as $E^{1}$ and
$E^{2}$, respectively, in this regime.
Indeed, in this limit the semiclassical computation is a good approximation, and as we found in Sec.~\ref{sec:semc}, the current responds as
$E^{\frac{D}{2}}$ in this regime.
\par
To compare the quantum vacuum expectation value of the induced current~(\ref{reg}) with the semiclassical current~(\ref{semi}), in $D=3$ dimension,
we plot $J_{\sem}$ as a function of the electric field in Fig.~\ref{fig1}.
This figure illustrates that in the strong electric regime $\lambda\gg\max(1,\lambdam)$, the semiclassical current approaches the induced current.
However, as one decreases $\lambda$ and goes away from the strong electric regime, there is a large discrepancy between the semiclassical current
and the induced current due to the exponential mass suppression factor in $J_{\sem}$;
see Eq.~(\ref{seminr}) and the discussion in Sec.~\ref{sec:heavy}.
%%%%%%%%%%%%%%%%%%%%%%%%%%%%%%%%%%%%%%%%%%%%%%%%%%%%%%%%%%%%%%%%%%%%%%%%%%%%%%%%%%%%%%%%%%%%%%%%%%%%%%%%%%%%%%%%%%%%%%%%%%%%%%%%%%%%%%%%%%%%%%%%%%%%%%%%%%%%%%%
%%%%%%%%%%%%%%%%%%%%%%%%%%%%%%%%%%%%%%%%%%%%%%%%%%%%%%%%%%%%%%%%%%%%%%%%%%%%%%%%%%%%%%%%%%%%%%%%%%%%%%%%%%%%%%%%%%%%%%%%%%%%%%%%%%%%%%%%%%%%%%%%%%%%%%%%%%%%%%%
\subsubsection{\label{sec:weak}Weak electric field regime}
The behavior of the current~(\ref{reg}) in the weak electric field regime $\lambda\ll\min(1,\lambdam)$ is obtained by a series expansion around
$\lambda=0$ with $\lambdam$ fixed. In the case of heavy particles, i.e., $\lambdam\gg 1$, the leading order terms are
\begin{align}\label{wheavy}
J&\simeq\frac{e^{2}H}{24\pi\mds}E, & \sigma&\simeq\frac{e^{2}H}{24\pi\mds},
\end{align}
and in the case of light particles, i.e., $\lambdam\ll 1$, leading order terms are given by
\begin{align}\label{wlight}
J&\simeq\frac{e^{2}H^{2}}{\pi^{2}\mds^{2}}E, & \sigma&\simeq\frac{e^{2}H^{2}}{\pi^{2}\mds^{2}}.
\end{align}
The results given by Eqs.~(\ref{wheavy}) and~(\ref{wlight}) are in agreement with the curves shown in Figs.~\ref{fig1} and~\ref{fig2}.
As illustrated in Fig.~\ref{fig1} the current monotonically increases for increasing electric field $E$.
Figure~\ref{fig2} shows that the conductivity is independent of the electric field $E$.
For both the current and the conductivity, the general feature is an inverse dependence on the scalar field mass parameter $\mds$.
In the case of $\ds_{2}$ the authors \cite{Frob:2014zka} showed that the current responds as $J\propto mE\exp(-2\pi m/H)$ for heavy particles and
behaves as $J\propto E/m^{2}$ for light particles, in this regime.
In the case of $\ds_{4}$, it has been shown that for the cases of heavy and light particles, the current behaves as $J\propto E/m^{2}$ \cite{Kobayashi:2014zza}.
%%%%%%%%%%%%%%%%%%%%%%%%%%%%%%%%%%%%%%%%%%%%%%%%%%%%%%%%%%%%%%%%%%%%%%%%%%%%%%%%%%%%%%%%%%%%%%%%%%%%%%%%%%%%%%%%%%%%%%%%%%%%%%%%%%%%%%%%%%%%%%%%%%%%%%%%%%%%%%%
%%%%%%%%%%%%%%%%%%%%%%%%%%%%%%%%%%%%%%%%%%%%%%%%%%%%%%%%%%%%%%%%%%%%%%%%%%%%%%%%%%%%%%%%%%%%%%%%%%%%%%%%%%%%%%%%%%%%%%%%%%%%%%%%%%%%%%%%%%%%%%%%%%%%%%%%%%%%%%%
\subsubsection{\label{sec:heavy}Heavy scalar field regime}
\begin{figure}
\includegraphics[scale=0.9]{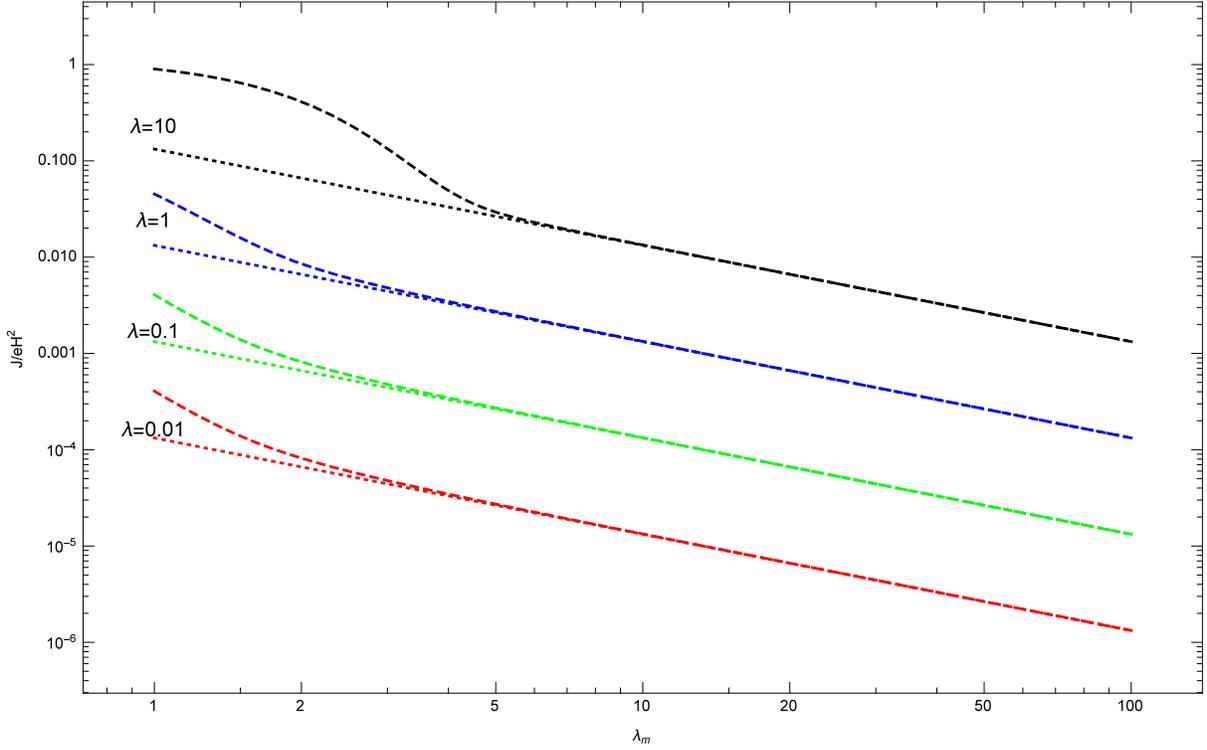}
\caption{For different values of $\lambda$, the normalized current~(\ref{reg}) and its approximation~(\ref{heavy}) are plotted as a function of
$\lambdam$ with dashed and dotted lines, respectively.} \label{fig3}
\end{figure}

The behavior of the current~(\ref{reg}) in the heavy scalar field regime $\lambdam\gg\max(1,\lambda)$ is obtained by taking the limit $\lambdam\rightarrow\infty$ with $\lambda$ held fixed. We then obtain the leading order terms as
\begin{align}\label{heavy}
J&\simeq\frac{e^{2}H}{24\pi\mds}E, & \sigma&\simeq\frac{e^{2}H}{24\pi\mds},
\end{align}
which are the same as the result~(\ref{wheavy}).
In Fig.~\ref{fig3}, the current~(\ref{reg}) and its limiting form~(\ref{heavy}) are plotted together.
This figure illustrates a good agreement between the exact and the approximated results.
\par
The semiclassical current~(\ref{seminr}), as it is exponentially suppressed, cannot appear in the plot of Fig.~\ref{fig3} and exponentially
disagrees with the current~(\ref{reg}).
The comparison of the results, in this regime, shows that in the $D=2$ dimension for heavy scalar \cite{Frob:2014zka} and for heavy fermion
\cite{Stahl:2015gaa} fields, the current scales as $J\propto m\sinh(2\pi\lambda)\exp(-2\pi m/H)$, i.e., is suppressed exponentially.
In $D=4$ dimension, for heavy scalar \cite{Kobayashi:2014zza} and for heavy fermion \cite{Hayashinaka:2016qqn} fields, the current scales as
$J\propto E/m^{2}$.
Therefore, in the heavy scalar field regime, due to an exponentially mass suppression factor, the semiclassical current does not agree with the
induced current.
%%%%%%%%%%%%%%%%%%%%%%%%%%%%%%%%%%%%%%%%%%%%%%%%%%%%%%%%%%%%%%%%%%%%%%%%%%%%%%%%%%%%%%%%%%%%%%%%%%%%%%%%%%%%%%%%%%%%%%%%%%%%%%%%%%%%%%%%%%%%%%%%%%%%%%%%%%%%%%%
%%%%%%%%%%%%%%%%%%%%%%%%%%%%%%%%%%%%%%%%%%%%%%%%%%%%%%%%%%%%%%%%%%%%%%%%%%%%%%%%%%%%%%%%%%%%%%%%%%%%%%%%%%%%%%%%%%%%%%%%%%%%%%%%%%%%%%%%%%%%%%%%%%%%%%%%%%%%%%%
\subsubsection{\label{sec:hyper}Massless minimally coupled scalar field case}
In the case of a massless minimally coupled scalar field, i.e., $\lambdam=0$, we now examine the behavior of the current in two limiting regimes.
In the limit $\lambda\rightarrow\infty$, the current and conductivity are approximated by Eq.~(\ref{strong}), whereas in the limit
$\lambda\rightarrow 0$ we find the leading order terms
\begin{align}\label{hyper}
J&\simeq\frac{H^{4}}{\pi^{2}E}, & \sigma&\simeq\frac{H^{4}}{\pi^{2}E^{2}}.
\end{align}
The results~(\ref{hyper}) agree with the asymptotic behavior of the red curves corresponding to $\lambdam=0$ in Figs.~\ref{fig1} and~\ref{fig2}.
In the regime $\lambda\ll 1$ the current and the conductivity are not bounded from above and increase as $\lambda$ decreases, as illustrated in Figs.~\ref{fig1} and~\ref{fig2}, respectively.
This divergence signals that the framework used to derive this result is not valid anymore and backreaction to the reservoir fields needs to be
taken into account. More about that regime will be given in Sec.~\ref{sec:IRHC}.
%%%%%%%%%%%%%%%%%%%%%%%%%%%%%%%%%%%%%%%%%%%%%%%%%%%%%%%%%%%%%%%%%%%%%%%%%%%%%%%%%%%%%%%%%%%%%%%%%%%%%%%%%%%%%%%%%%%%%%%%%%%%%%%%%%%%%%%%%%%%%%%%%%%%%%%%%%%%%%%
%%%%%%%%%%%%%%%%%%%%%%%%%%%%%%%%%%%%%%%%%%%%%%%%%%%%%%%%%%%%%%%%%%%%%%%%%%%%%%%%%%%%%%%%%%%%%%%%%%%%%%%%%%%%%%%%%%%%%%%%%%%%%%%%%%%%%%%%%%%%%%%%%%%%%%%%%%%%%%%
\begin{figure}
\includegraphics[scale=0.9]{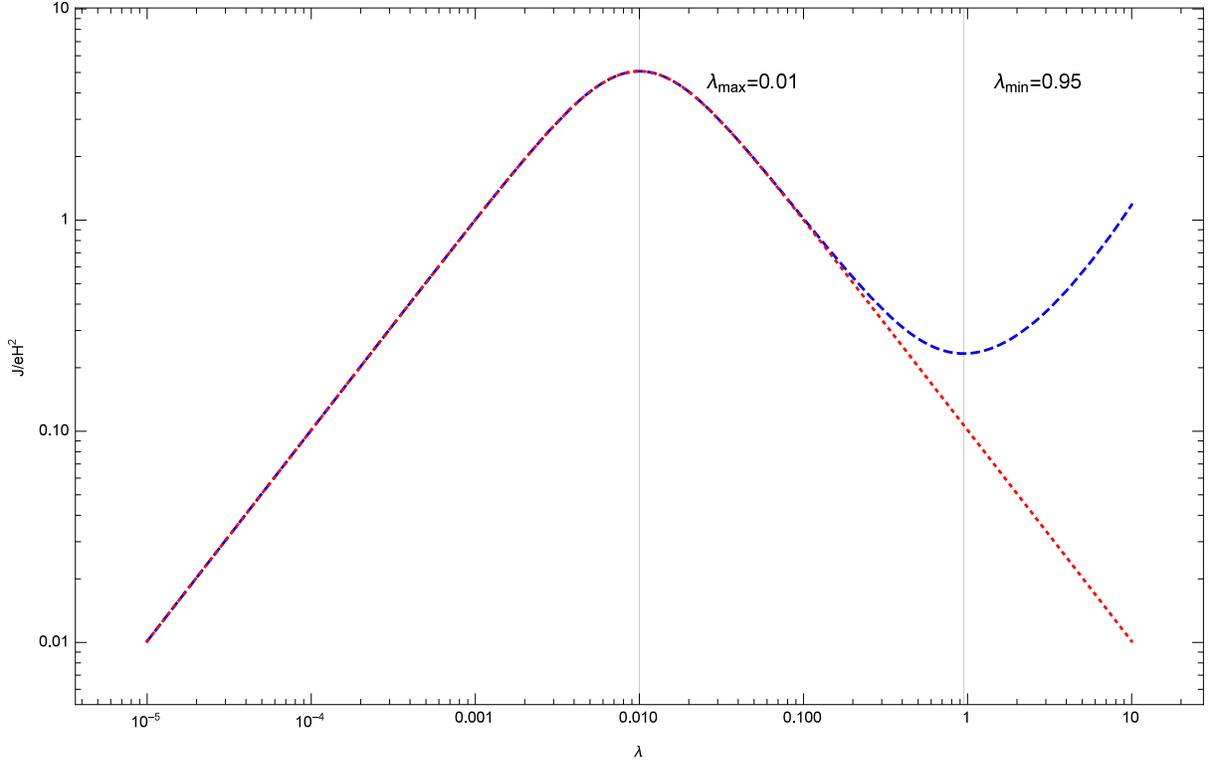}
\caption{The normalized current~(\ref{reg}) (in the blue dashed line) and its approximation~(\ref{light}) (in the red dotted line) are plotted
as a function of $\lambda$ with $\lambdam=0.01$.} \label{fig4}
\end{figure}
\subsubsection{\label{sec:light}Light scalar field case}
We now study the behavior of the current~(\ref{reg}) for a light scalar field case, i.e., $\lambdam\ll 1$ and more specifically
$\lambdam<\sqrt{3/4}$. In the regime $\lambda\gg 1$, the current and conductivity scale as indicated by Eq.~(\ref{strong}).
Numerical analyses show that in the regime $0\lesssim\lambda\lesssim 1$ the current and conductivity behave as
\begin{align}\label{light}
J&\simeq\frac{eH^{2}}{\pi^{2}}\Big(\frac{\lambda}{\lambda^{2}+\lambdam^{2}}\Big), &
\sigma&\simeq\frac{e^{2}}{\pi^{2}}\Big(\frac{1}{\lambda^{2}+\lambdam^{2}}\Big).
\end{align}
In Sec.~\ref{sec:IRHC} we will derive Eq.~(\ref{light}) analytically. In Fig.~\ref{fig4}, we plot the current~(\ref{light}) together with the current~(\ref{reg}), in the IR-HC regime.
This figure illustrates the quite good agreement between the numerical and analytical results.
The current has a local minimum in $\lambda_{\mathrm{min}}\simeq\sqrt{3/4}+\epsilon$, where $\epsilon$ is a small positive parameter, and a local
maximum in $\lambda_{\mathrm{max}}\simeq\lambdam$.
In the interval $\lambda\in\big(\lambda_{\mathrm{max}},\lambda_{\mathrm{min}}\big)$ the phenomenon of IR-HC occurs; i.e., the current increases
for decreasing $\lambda$.
Beyond $\lambda_{\mathrm{max}}$, in the interval $(0,\lambda_{\mathrm{max}})$, the current has a linear response for $\lambdam\neq 0$ which agrees
with Eq.~(\ref{wlight}).
From this and Sec.~\ref{sec:hyper}, we can conclude that for decreasing $\lambda$, if $\gamma$ becomes real, then there would be a period of IR-HC;
if one keeps decreasing $\lambda$, it will be followed by a linear behavior for $\lambdam\neq 0$ or continued unbounded for $\lambdam=0$.
\par
In this section, we computed the current and described it in different limiting cases.
Especially a IR-HC regime has been reported, and we propose a discussion and a summary of what is known about IR-HC in Sec.~\ref{sec:IRHC}.
%%%%%%%%%%%%%%%%%%%%%%%%%%%%%%%%%%%%%%%%%%%%%%%%%%%%%%%%%%%%%%%%%%%%%%%%%%%%%%%%%%%%%%%%%%%%%%%%%%%%%%%%%%%%%%%%%%%%%%%%%%%%%%%%%%%%%%%%%%%%%%%%%%%%%%%%%%%%%%%
%%%%%%%%%%%%%%%%%%%%%%%%%%%%%%%%%%%%%%%%%%%%%%%%%%%%%%%%%%%%%%%%%%%%%%%%%%%%%%%%%%%%%%%%%%%%%%%%%%%%%%%%%%%%%%%%%%%%%%%%%%%%%%%%%%%%%%%%%%%%%%%%%%%%%%%%%%%%%%%
\section{\label{sec:IRHC}Discussion about IR-HC in $D$ dimensions}
IR-HC is a regime where for a given interval of the electric field, a decreasing electric field gives an increasing conductivity.
IR-HC was first reported in \cite{Frob:2014zka}.
The authors showed there, in the case of $\ds_{2}$ that the current responds as $J\sim E^{-1}$ for small electric fields and IR-HC was present
for $m/H<1/2$. In the case of $\ds_{4}$ the renormalization scheme introduces a term of the form $\log(m/H)$ \cite{Kobayashi:2014zza} in the
regularized current expression which arises from the second order adiabatic expansion.
Therefore, it signals this renormalization method was not applicable for the case of exactly massless scalar field in $\ds_{4}$
\cite{Kobayashi:2014zza}; see also discussions in \cite{Parker:1974qw,Fulling:1974zr}.
Hence it was not possible to discuss IR-HC for the massless case but it was present for $m/H<\sqrt{5/4}$. In \cite{Hayashinaka:2016dnt}, the
point-splitting method was shown to agree with the adiabatic subtraction in $\ds_{4}$ for boson.
In $\ds_{3}$, we report the same behavior $J\sim E^{-1}$ in the regime of small electric field and massless minimally coupled charged particles
and report IR-HC for $\mds/H<\sqrt{3/4}$.
\par
These results lead us to propose a procedure to avoid an IR-HC regime by setting the value of the conformal coupling to a specific range.
In $\dsd$, the \textit{nonrenormalized} in-vacuum state expectation of the spacelike component of the current operator is
\begin{align}\label{vevd}
\langle j^{1}\rangle_{in}&=\,_{in}\langle0|j^{1}|0\rangle_{in} \nn\\
&=2e\Omega^{-D}(\tau)\int\frac{d^{d}k}{(2\pi)^{d}}\big(k_{x}+eA_{1}(\tau)\big)
\frac{e^{\kappa\pi i}}{2k}\big|\wwp(z_{+})\big|^{2}.
\end{align}
Generalizing to $D$ dimensions the step performed between Eq.~(\ref{vev}) and Eq.~(\ref{curent}), the integral~(\ref{vevd}) can be conveniently
rewritten as
\begin{equation}
\langle j^{1}\rangle_{in}=\frac{eH^{d}}{(2\pi)^{d}}\Omega^{-1}(\tau)\int d\Sigma_{d-1}e^{\lambda r\pi}
\lim_{\Lambda\rightarrow\infty}
\int_{0}^{\Lambda}dp\,p^{d-2}\big(rp-\lambda\big)\big|\wwp(z_{+})\big|^{2}, \label{integrald}
\end{equation}
where $d\Sigma_{d-1}$ is given by Eq.~(\ref{aelement}).
As pointed out in \cite{Frob:2014zka,Kobayashi:2014zza}, in a IR-HC regime, the population of created pairs is dominated by IR contribution and
no longer by the pairs created within a Hubble time.
Hence the asymptotic behavior of the wave function, in the limit $p\rightarrow 0$ will give the dominant term in a IR-HC regime.
The Whittaker function $\wwp(z)$ as $z\rightarrow 0$, has an asymptotic form \cite{book:Nist} given by
\begin{equation}\label{irw}
\wwp(z)\sim z^{\frac{1}{2}-\gamma},
\end{equation}
and in this regime $\gamma$ is real and by convention is positive.
As a consequence, the integral~(\ref{integrald}) in the limit $p\rightarrow0$ behaves as
\begin{equation}\label{irint}
\lambda\int_{0}dp\,p^{D-2-2\gamma}.
\end{equation}
Then power counting shows that in the regime
\begin{equation}\label{ircon}
\gamma>\frac{D-2}{2},
\end{equation}
the current integrand diverges in the limit $p\rightarrow 0$.
However, since $\gamma\leq\frac{D-1}{2}$, the total current integral remains finite.
From Eq.~(\ref{ircon}) and the definition of $\gamma$, given by Eq.~(\ref{gamma}), we find first that
\begin{align}
\label{eq:approx}
\langle j^{1}\rangle_{in}\propto\frac{eH^{d}\lambda}{\rho^2}
=\frac{eH^{d}\lambda}{\lambda^{2}+\lambdam^{2}}, && \rho\ll d.
\end{align}
Observe that setting $\lambdam=0$, one recovers the behavior $J\propto E^{-1}$.
Figure~\ref{fig4} shows a plot of the current in the IR-HC regime together with the analytical result of~(\ref{eq:approx}), and the curves agree
reasonably well. Similar plots could be produced for $D=2$ or $D=4$.
Second, again from Eqs.~(\ref{gamma}) and~(\ref{ircon}) it is also possible to deduce that IR-HC occurred when
\begin{equation}\label{iregime}
\lambdam^{2}+\lambda^{2}<\frac{2D-3}{4}.
\end{equation}
Therefore, a sufficient condition to avoid IR-HC is
\begin{equation}\label{avoid}
\lambdam^{2}\geq\frac{2D-3}{4},
\end{equation}
and we define thus $\lambda_{\text{m,min}}=\frac{\sqrt{2D-3}}{2}$. The previous condition implies also a minimal bound to avoid IR-HC for the
conformal coupling $\xi$ which in the case of a massless scalar field reads
\begin{equation}\label{ximin}
\xi_{\mathrm{min}}=\frac{2D-3}{4D(D-1)}.
\end{equation}
We see that in nonconformally coupled theories, a conformal coupling with values larger than $\xi_{\text{min}}$ can be used to avoid the IR-HC
regime. Conversely, $\forall\xi<\xi_{\mathrm{min}}$, IR-HC would appear for $m^{2}/H^{2}\in\mathcal{I}_{\text{IR-HC}}$, with
\begin{equation}\label{mathi}
\mathcal{I}_{\text{IR-HC}}:=\big(0,\frac{2D-3}{4}-D(D-1)\xi\big).
\end{equation}
These results are summarized in Table~\ref{tableIRHC}.
\begin{table}
\begin{tabular}{|c|c|c|c|}
  \hline
  & $\lambda_{\text{m,min}}^{2}$ & $\xi_{\mathrm{min}}$ & $\mathcal{I}_{\text{IR-HC}}$  \\
  \hline
  In $\ds_{2}$ & $\frac{1}{4}$    & $\frac{1}{8}$ &  $\big(0,\frac{1}{4}-2\xi\big)$ \\
  In $\ds_{3}$ & $\frac{3}{4}$    & $\frac{1}{8}$ &  $\big(0,\frac{3}{4}-6\xi\big)$  \\
  In $\ds_{4}$ & $\frac{5}{4}$    & $\frac{5}{48}$ & $\big(0,\frac{5}{4}-12\xi\big)$  \\
  In $\dsd$ & $\frac{2D-3}{4}$ & $\frac{2D-3}{4D(D-1)}$ & $\big(0,\frac{2D-3}{4}-D(D-1)\xi\big)$  \\
  \hline
\end{tabular}
\caption{Minimum values of $\lambdam^{2}$ and $\xi$ to avoid IR-HC.
The interval $\mathcal{I}_{\text{IR-HC}}$ is the range of $m^{2}/H^{2}$ for which IR-HC would appear after turning on the conformal coupling.} \label{tableIRHC}
\end{table}
In $D=2$ and $D=3$, Eq.~(\ref{avoid}) agrees very well with numerical investigations.
However, in the case of $D=4$, to avoid IR-HC one needs to have $\lambdam\gtrsim 1.25$ \cite{Kobayashi:2014zza}; in this case, a small variation
from condition~(\ref{avoid}), comes from a term also dominant in the IR regime but not taken into account in the previous calculation: the one
coming from the renormalization in "$\log(m/H)$."
\par
Table~\ref{tablelminlmax} presents the results of the numerical investigations for the value of $\lambda_{\text{min}}$ and $\lambda_{\text{max}}$
for dimensions $D=2,3,4$.
Numerical investigations indicate that $\lambda_{\text{max}}=\lambdam$ and $\lambda_{\text{min}}=\lambda_{\text{m,min}}+\epsilon$ with $\epsilon>0$.
Recall that $\lambda_{\text{m,min}}=0.5,0.87,1.12$ for $D=2,3,4$, respectively.
$\epsilon$ reaches an asymptotic value for $\lambdam\rightarrow 0$ in $D=2,3$, whereas it is unbounded for $D=4$.
This difference comes again from the renormalization term in "$\log(m/H)$."
\par
Looking at the fermionic induced current in $\ds_{2}$ \cite{Stahl:2015gaa} and $\ds_{4}$ \cite{Hayashinaka:2016qqn}, no IR-HC was reported.
In $\ds_{2}$, the only difference between the fermion and the boson was effectively a translation of the mass squared, i.e., $m^2_{\mathrm{fermion}}=m^2_{\mathrm{boson}}-H^2/4$.
It is furthermore known that a massless fermion is conformally invariant and gives, as in flat spacetime, a linear behavior for the current.
In the bosonic case this conformal behavior was found for $m^{2}/H^{2}=1/4$ and the IR-HC for $0\leq m^{2}/H^{2}<1/4$.
Hence, conformality plays an important role to understand IR-HC.
Note that for a fermionic particle in $D=2$, to have a regime of IR-HC one needs to let the mass parameters $m^{2}/H^{2}<0$, that is, to allow
for tachyonic propagation.
In parallel to tachyon, IR-HC is a regime where decreasing one source (the electrical field) increases the consequence (the produced pairs).
Therefore, it is against physical intuition and for massless cases leads even to a current unbounded from above.
The links between tachyonic field, conformality, and IR-HC remain to be explored.
\begin{table}
\begin{tabular}{|c|c c c|c c c|c c c|}
  \hline
  $D$  &  & 2 & &  & 3 & & & 4 &  \\
  \hline
  $\lambdam$ & 0.1,&0.01,&0.001 &0.1,&0.01,&0.001 & 0.1,&0.01,&0.001 \\
  \hline
  $\lambda_{\text{min}}$ & 0.54,&0.59,&0.59 &0.92,&0.95,&0.95 &1.27,&1.54,&1.68 \\
  \hline
  $\lambda_{\text{max}}$ &0.1,&0.01,&0.001 &0.1,&0.01,&0.001 &0.1,&0.01,&0.001 \\
  \hline
\end{tabular}
\caption{Numerically found values of $\lambda_{\text{min}}$ and $\lambda_{\text{max}}$ for different values of $\lambdam$ and $D$.
IR-HC occurs for $\lambda\in(\lambda_{\text{max}},\lambda_{\text{min}})$.
Those results point toward the idea that $\lambda_{\text{max}}=\lambdam$ and $\lambda_{\text{min}}=\lambda_{\text{m,min}}+\epsilon=\frac{\sqrt{2D-3}}{2}+\epsilon$, $\epsilon>0$.} \label{tablelminlmax}
\end{table}
%%%%%%%%%%%%%%%%%%%%%%%%%%%%%%%%%%%%%%%%%%%%%%%%%%%%%%%%%%%%%%%%%%%%%%%%%%%%%%%%%%%%%%%%%%%%%%%%%%%%%%%%%%%%%%%%%%%%%%%%%%%%%%%%%%%%%%%%%%%%%%%%%%%%%%%%%%%%%%%
%%%%%%%%%%%%%%%%%%%%%%%%%%%%%%%%%%%%%%%%%%%%%%%%%%%%%%%%%%%%%%%%%%%%%%%%%%%%%%%%%%%%%%%%%%%%%%%%%%%%%%%%%%%%%%%%%%%%%%%%%%%%%%%%%%%%%%%%%%%%%%%%%%%%%%%%%%%%%%%
\section{\label{sec:gravity}discussion about gravitational backreaction}
In this last section, we present our first results on the gravitational backreaction.
More specifically, our main goal is to naively estimate the variation of the Hubble constant in the heavy scalar field regime.
The numerous works on the Schwinger effect in $\ds$
\cite{Frob:2014zka,Kobayashi:2014zza,Fischler:2014ama,Stahl:2015gaa,Hayashinaka:2016qqn,Hayashinaka:2016dnt,Yokoyama:2015wws}
always assumed that the created pairs do not backreact to the background metric.
This assumption holds as far as the energy density of the pairs is much smaller than the background Hubble energy \cite{Xue:2014kna,Xue:2015tmw}.
For this paper, we will focus on a semiclassical computation of the stress energy-momentum tensor.
We assume that the effects of the pair creation to the Einstein equation are small; they give rise to an effective cosmological constant $\Lambda_{\mathrm{eff}}$ in the Einstein equation.
Then, the Einstein equation can be written as
\begin{equation}\label{einstein}
R^{\mu\nu}-\frac{1}{2}Rg^{\mu\nu}+\Lambda_{\mathrm{eff}}g^{\mu\nu}=-8\pi G_{D}T^{\mu\nu}_{\sem},
\end{equation}
where $R^{\mu\nu}$ is the $\dsd$ Ricci tensor and $G_{D}=H^{4-D}M_{\mathrm{P}}^{-2}$ \cite{Book:Zwiebach} is the gravitational constant in $D$
dimensions, with $M_{\mathrm{P}}$ being the Planck mass.
Now, we wish to compute the semiclassical energy-momentum tensor on the right-hand side of the Einstein equation~(\ref{einstein}).
Similar to, e.g., \cite{Book:Padmanabhan} the semiclassical energy-momentum tensor of the Schwinger pairs can be defined as
\begin{equation}\label{emtdef}
T^{\mu\nu}_{\sem}:=
|g|^{\frac{-1}{2}}\int\frac{d^{d}k}{(2\pi)^{d}}\frac{p^{\mu}_{\k}p^{\nu}_{\k}}{p^{0}_{\k}}|\beta_{\k}|^{2},
\end{equation}
where $|\beta_{\k}|^{2}$~(\ref{betasq}) is the distribution function and $p^{\mu}_{\k}$~(\ref{momentum}) is the physical momentum vector of the
created particle.
To perform the integral on the right-hand side of Eq.~(\ref{emtdef}), we follow the same integration procedure used in Sec.~\ref{sec:den}: impose
the relation~(\ref{extremum}) to convert the $k$ integral into a $\tau$ integral.
In the heavy scalar field regime, $\lambdam\gg\max(1,\lambda)$, we consider a terminal value of the physical momentum.
Hence, substituting expressions~(\ref{terminalm}) and~(\ref{terminale}) into Eq.~(\ref{emtdef}) leads to
\begin{eqnarray}\label{emt}
T^{00}_{\sem}&\simeq&\Omega^{-2}(\tau)\mathcal{E}, \hspace{1cm}
T^{01}_{\sem}\simeq-\frac{\lambda}{\lambdam}T^{00}_{\sem}, \hspace{1cm}
T^{11}_{\sem}\simeq\frac{\lambda^{2}}{\lambdam^{2}}T^{00}_{\sem}, \nn\\
T^{0i}_{\sem}&=&T^{ij}_{\sem}=0, \hspace{1cm} i=2,\ldots,d,
\end{eqnarray}
where $\mathcal{E}$ is given by
\begin{equation}\label{mathcale}
\mathcal{E}=\frac{H^{D}}{(2\pi)^{D-2}(D-1)}\lambda^{\frac{3-D}{2}}\I_{\frac{D-3}{2}}(2\pi\lambda)\lambdam^{D}e^{-2\pi\lambdam}.
\end{equation}
\par
The Hubble parameter is defined as
\begin{equation}\label{parameter}
H(\tau):=\Omega^{-2}(\tau)\frac{d\Omega(\tau)}{d\tau}.
\end{equation}
Considering the metric~(\ref{metric}), in terms of the Hubble parameter $H(\tau)$, the components of the Ricci tensor are obtained
\begin{eqnarray}\label{riccitensor}
R_{00}&=&(D-1)\Big(H^{2}(\tau)+\Omega^{-1}(\tau)\dot{H}(\tau)\Big)\Omega^{2}(\tau), \nn \\
R_{ij}&=&-\Big((D-1)H^{2}(\tau)+\Omega^{-1}(\tau)\dot{H}(\tau)\Big)\Omega^{2}(\tau)\delta_{ij}, \nn \\
R_{0i}&=&0, \hspace{1cm} i=1,\ldots,d,
\end{eqnarray}
and the Ricci scalar is
\begin{equation}\label{ricciscalar}
R=(D-1)\Big(DH^{2}(\tau)+2\Omega^{-1}(\tau)\dot{H}(\tau)\Big).
\end{equation}
The trace of the Einstein equation~(\ref{einstein}) gives
\begin{equation}\label{Lambda}
\Lambda_{\mathrm{eff}}=\frac{(D-2)R}{2D}-\frac{8\pi G_{D}\mathcal{E}}{D},
\end{equation}
and in the heavy scalar field regime, we find that the leading order terms for the Einstein equation~(\ref{einstein}) involve $T_{\sem}^{00}$:
using Eqs.~(\ref{riccitensor})-(\ref{Lambda}) it leads to
\begin{equation}\label{einsteineq}
\Omega^{-1}(\tau)\frac{dH(\tau)}{d\tau}=-\frac{8\pi G_{D}\mathcal{E}}{(D-2)}.
\end{equation}
The above equation determines the evolution of the Hubble parameter with respect to the conformal time $\tau$.
In order to compare with the existing literature, we now work in cosmic time $t$: using Eqs.~(\ref{line}) and~(\ref{metric}), it can be shown
that the evolution of the Hubble parameter with respect to the cosmic time $t$ is
\begin{equation}\label{hubbleq}
\frac{dH(t)}{dt}=-\frac{8\pi G_{D}\mathcal{E}}{(D-2)},
\end{equation}
which agrees with \cite{Mottola:1984ar,Markkanen:2016aes,Cai:2005ra}.
Thus, the Schwinger pair creation leads to a decay of the Hubble constant and, as consequence of Eq.~(\ref{Lambda}), a decay of the cosmological
constant. This decay of the cosmological constant begins with the pair creation and continues until $\Lambda_{\mathrm{eff}}=0$.
In this picture, as a classical black hole being evaporated into Hawking radiation or the coherent energy of an electric field being dissipated
into $e^+$ $e^-$ pairs, the coherent vacuum energy is dissipated into a cloud of scalar pairs.
The decay of the Hubble constant affects $G_{D}$ for $D\neq 4$.
For $D<4$ the gravitational constant decays until it reaches zero and for $D>4$ the gravitational constant increases.
Similar to \cite{Mottola:1984ar}, the time scale for evolution of the Hubble constant can be estimated by
\begin{equation}\label{timescale}
t_{B}:=-\frac{H}{\frac{dH(t)}{dt}}=\frac{(2\pi)^{D-3}(D-1)(D-2)M_{\mathrm{P}}^{2}}{4H^{3}}
\Big(\lambda^{\frac{3-D}{2}}\I_{\frac{D-3}{2}}(2\pi\lambda)\Big)^{-1}\lambdam^{-D}e^{2\pi\lambdam}.
\end{equation}
A series expansion of the time scale expression~(\ref{timescale}) around $\lambda=0$, with $\lambdam$ fixed, leads to the leading order term
\begin{equation}\label{behavior}
t_{B}\simeq\frac{(4\pi)^{\frac{D-3}{2}}\Gamma\big(\frac{D-1}{2}\big)(D-1)(D-2)M_{\mathrm{P}}^{2}}{4H^{3}}\lambdam^{-D}e^{2\pi\lambdam},
\end{equation}
which is independent of $\lambda$.
In \cite{Mottola:1984ar}, the time scale has been computed in the global patch of $\ds_{4}$, without electric field, and the author showed there,
in the limit $m \gg H$, the time scale behaves as $Hm^{-4}\exp(\pi m/H)$.
Hence, in $D=4$ dimension, the result~(\ref{behavior}) agrees with the time scale obtained in the Ref.~\cite{Mottola:1984ar} up to a factor of 2
in the exponent. This factor could come from the different definitions for the energy-momentum tensor.
\par
Observe that the calculation carried out in this section is not valid for $D =2$ as there is no Einstein gravity in 1+1 dimension.
Observe beside that under our working assumption, i.e., heavy scalar field regime, $\lambdam\gg\max(1,\lambda)$, we find
$t_{B} \gg t_{\text{H}}=H^{-1}$ which still allow for a long inflation.
Furthermore, we argue that this decay of the Hubble constant presents similarities with generic models of slow roll inflation where a scalar field
sees its potential energy slowly decaying into kinetic energy to ultimately exhibit coherent oscillations around the minimum of its potential which
unleash a reheating phase.
We want to explore further this issue in a future paper \cite{inprep}.
The next step is to consider the expectation value of the energy-momentum operator, which as the current will present divergences.
The computation of this tensor is much more involving and is beyond the scope of this paper.
We have seen in Sec.~\ref{sec:result} that the semiclassical estimates agreed in the strong field regime, but were exponentially different in
the heavy scalar regime, so we argue that those results have to be checked by further study, mainly the exact computation of the energy-momentum
tensor in order to see if those first estimates agree with the general case.
For instance, a very recent work \cite{Markkanen:2016aes}, without electric field, in $D=4$ dimension, with a slightly different method, discovered
an enhancement of the Hubble constant.
The same exponential behavior as in Eq.~(\ref{timescale}) was also found but with a different prefactor.
We argue that those changes are due to the renormalization procedure they carried out which gives different results than the replacement of the
$k$ integral into a $\tau$ integral we performed here.
%%%%%%%%%%%%%%%%%%%%%%%%%%%%%%%%%%%%%%%%%%%%%%%%%%%%%%%%%%%%%%%%%%%%%%%%%%%%%%%%%%%%%%%%%%%%%%%%%%%%%%%%%%%%%%%%%%%%%%%%%%%%%%%%%%%%%%%%%%%%%%%%%%%%%%%%%%%%%%%
%%%%%%%%%%%%%%%%%%%%%%%%%%%%%%%%%%%%%%%%%%%%%%%%%%%%%%%%%%%%%%%%%%%%%%%%%%%%%%%%%%%%%%%%%%%%%%%%%%%%%%%%%%%%%%%%%%%%%%%%%%%%%%%%%%%%%%%%%%%%%%%%%%%%%%%%%%%%%%%
\section{\label{sec:concl}Conclusion}
We have investigated pair creation by the Schwinger mechanism in $\dsd$.
Specifically, we considered a charged massive scalar field coupled to a constant background electric field in $\dsd$.
After the canonical quantization, Bogoliubov coefficients were obtained, and then the decay rate and the density of created pairs were computed;
see these main results in Eqs.~(\ref{rate}) and~(\ref{density}).
Also, using a semiclassical method the decay rate and the density were computed; see Appendix~\ref{app:sem}.
Both methods agree to say that in the semiclassical approximation, the screening orientation stays and the antiscreening ordination is suppressed.
The density of created pairs is constant with respect to time.
It signals that the pair creation in $\dsd$ from electric and gravitational fields exactly balances the dilution from the expansion of the universe.
Under the semiclassical condition we computed the conduction current of the created particles in any dimension.
We find that in the strong electric field regime, $\lambda\gg\max(1,\lambdam)$, the semiclassical current becomes independent of the scalar field
mass and responds as $E^{\frac{D}{2}}$, and in the heavy scalar field regime, $\lambdam\gg\max(1,\lambda)$, due to the presence of a Boltzmann
mass suppression factor it exponentially damped.
Our main goal has been to study the induced quantum vacuum expectation value of the conduction current of the created pairs.
Thus, in the case of a $D=3$ dimensional dS, the expectation value of the spacelike component of the current operator has been computed in the
in-vacuum state. As expected, a linear UV divergence appeared.
Applying an adiabatic subtraction regularization scheme the divergent term was removed and a finite expression was obtained for the current and
the corresponding conductivity.
They have been plotted in Figs.~\ref{fig1} and~\ref{fig2}, respectively.
The current and conductivity have been also analytically investigated.
We find that in the strong electric field regime, $\lambda\gg\max(1,\lambdam)$, the current responds as $E^{\frac{3}{2}}$ and becomes independent
of scalar field mass parameter $\mds$~(\ref{mds}).
In the weak electric field regime, $\lambda\ll\min(1,\lambdam)$, the current has a linear response in $E$ and is inversely proportional to $\mds$.
For the case of a massless minimally coupled scalar field, i.e., $\lambdam=0$, for $\lambda\lesssim 1$, the current varies as $E^{-1}$.
Consequently, in this regime, the current and conductivity are increasing unbounded for decreasing electric field, which leads to the phenomenon
of IR-HC. The regime of IR-HC has been extensively discussed in Sec.~\ref{sec:IRHC} from both numerical and analytical points of view.
It has been shown that IR-HC happens for $\lambda\in\big(\lambdam,\frac{\sqrt{2D-3}}{2}+\epsilon\big)$ with $0<\epsilon\ll 1$ for $D=2,3$ and
$\epsilon$ positive but unbounded in $D=4$.
This difference comes from the renormalization scheme used in $D=4$.
The behavior of the current has also been derived in the IR-HC regime for any dimension in Eq.~(\ref{eq:approx}) up to the renormalization factors.
A proposed relation of IR-HC with conformality and tachyonicity remains also to be further explored but is beyond the scope of this paper.
\par
Until Sec.~\ref{sec:gravity}, the gravitational and electric fields were treated as an external field, and one important next step is to take into
account backreactions of the created pairs to those two fields.
Indeed, as soon as the energy of the population of the Schwinger created pairs becomes of the order of the energy carried by the constant electric
field or of the gravitational energy, backreaction effects become unavoidable.
Investigating these effects could be used to find specific forms of electric fields or specific classes of spacetimes which favor or disfavor
pair creation.
Furthermore, it could also be a fruitful way to make cosmological statements about magnetogenesis, matter-antimatter asymmetry, primordial
gravitational waves, or the way inflation is driven and ends.
Those issues are currently under investigation \cite{inprep}.
Our first results on gravitational backreaction effects were depicted in Sec.~\ref{sec:gravity}.
Using a semiclassical approach the energy-momentum tensor of the Schwinger pairs has been computed in the heavy scalar field regime; see
Eq.~(\ref{emt}). We showed that creation of particles leads to a decay of the Hubble constant.
In the limit of zero electric field, our result is consistent with a previous study \cite{Mottola:1984ar} up to a factor of 2 in the exponent but
disagrees with a very recent work \cite{Markkanen:2016aes}.
A more consistent calculation of this effect must dynamically study the evolution of the Hubble constant $H$ through Einstein equations, and this
will explicitly break de Sitter invariance by introducing a preferred time slicing.
We argue that it should be possible to compute it together with the corrections from the Schwinger effect and the presence of an electric field
to the vacuum fluctuation during an inflationary phase.
This could in turn affect the power spectrum at the end of inflation, as it was already suggested in the conclusion of \cite{Pioline:2005pf}.
After the evolution of the primordial power spectrum through the reheating and the radiation dominated era, in principle it could be measured by
cosmic microwave background experiments.
%%%%%%%%%%%%%%%%%%%%%%%%%%%%%%%%%%%%%%%%%%%%%%%%%%%%%%%%%%%%%%%%%%%%%%%%%%%%%%%%%%%%%%%%%%%%%%%%%%%%%%%%%%%%%%%%%%%%%%%%%%%%%%%%%%%%%%%%%%%%%%%%%%%%%%%%%%%%%%%
%%%%%%%%%%%%%%%%%%%%%%%%%%%%%%%%%%%%%%%%%%%%%%%%%%%%%%%%%%%%%%%%%%%%%%%%%%%%%%%%%%%%%%%%%%%%%%%%%%%%%%%%%%%%%%%%%%%%%%%%%%%%%%%%%%%%%%%%%%%%%%%%%%%%%%%%%%%%%%%
\section*{Acknowledgements}
S.-S.~X. thank Professor R. Ruffini for discussions on general relativity, cosmology and physics of pair productions.
C.~S. is supported by the Erasmus Mundus Joint Doctorate Program by Grant No.~2013-1471 from the EACEA of the European Commission.
E.~B. is supported by the University of Kashan Grant No.~317203/1.
%%%%%%%%%%%%%%%%%%%%%%%%%%%%%%%%%%%%%%%%%%%%%%%%%%%%%%%%%%%%%%%%%%%%%%%%%%%%%%%%%%%%%%%%%%%%%%%%%%%%%%%%%%%%%%%%%%%%%%%%%%%%%%%%%%%%%%%%%%%%%%%%%%%%%%%%%%%%%%%
%%%%%%%%%%%%%%%%%%%%%%%%%%%%%%%%%%%%%%%%%%%%%%%%%%%%%%%%%%%%%%%%%%%%%%%%%%%%%%%%%%%%%%%%%%%%%%%%%%%%%%%%%%%%%%%%%%%%%%%%%%%%%%%%%%%%%%%%%%%%%%%%%%%%%%%%%%%%%%%
\appendix
\section{\label{app:sem}Semiclassical scattering method}
In this appendix we compute the pair creation rate using the semiclassical scattering method. Starting from Eq.~(\ref{refeq}), it is possible to
write the equation of motion for the scalar particles as a harmonic oscillator one with a time dependent frequency given by Eq.~(\ref{omega}).
From here it is possible to do a Bogoliubov transformation and reformulate the equation of motion~(\ref{refeq}) in term of the Bogoliubov
coefficients $\alpha_{\k}(\tau)$ and $\beta_{\k}(\tau)$.
This technique is inspired from well known flat spacetime techniques (see, e.g., \cite{Dumlu:2011rr}) and was already applied to the equivalent
problem in $D=4$ dimension \cite{Stahl:2015cra}. The result will be similar up to dimensional factors.
The semiclassical scattering method was usually referred to as the WKB method, but as detailed in \cite{Blinne:2015zpa}, it is more precise to call
it the scattering semiclassical method to differentiate with other WKB inspired methods \cite{Kleinert:2008sj}.
The starting point is to implement the Bogoliubov transformation using an ansatz inspired by a WKB expansion,
\begin{align}
\label{wkbf}
f_{\k}(\tau)&=\frac{\alpha_{\k}(\tau)}{\sqrt{\omega(\tau)}}e^{-iK(\tau)}
+\frac{\beta_{\k}(\tau)}{\sqrt{\omega(\tau)}}e^{iK(\tau)}, \\
\label{wkbfd}
\dot{f}_{\k}(\tau)&=-i\omega(\tau)\bigg[\frac{\alpha_{\k}(\tau)}{\sqrt{\omega(\tau)}}e^{-iK(\tau)}
-\frac{\beta_{\k}(\tau)}{\sqrt{\omega(\tau)}}e^{iK(\tau)}\bigg],
\end{align}
where
\begin{equation}\label{keq}
K(\tau)=\int_{-\infty}^{\tau}\omega(\tau')d\tau'.
\end{equation}
To preserve the commutation relation, it is necessary to impose the Wronskian condition $|\alpha_{\k}(\tau)|^2-|\beta_{\k}(\tau)|^2=1$.
In this basis, the momentum spectrum of the pair creation rate reads
\begin{equation}\label{spectrum}
n_{\k}=\lim_{\tau\rightarrow0}\left|\beta_{\k}(\tau)\right|^2,
\end{equation}
with the boundary conditions being plane waves in the positive frequency direction as the past asymptotic behavior of Eq.~(\ref{refeq}) suggests
\begin{align}\label{boundary}
\beta_{\k}(-\infty)=0, && \alpha_{\k}(-\infty)=1.
\end{align}
It is possible to find a first order coupled differential equation for the Bogoliubov coefficients
\begin{align}
\label{alphadot}
\dot{\alpha}_{\k}(\tau)&=\frac{\dot{\omega}(\tau)}{2\omega(\tau)}e^{2iK(\tau)}\beta_{\k}(\tau), \\
\label{betadot}
\dot{\beta}_{\k}(\tau)&=\frac{\dot{\omega}(\tau)}{2\omega(\tau)}e^{-2iK(\tau)}\alpha_{\k}(\tau).
\end{align}
Note that at this point, the equations derived are still exact.
Aiming at finding the momentum spectrum~(\ref{spectrum}), it is possible to integrate formally Eqs.~(\ref{alphadot}) and~(\ref{betadot}) by using
the boundary condition~(\ref{boundary}). One finds
\cite{Berry:1982}
\begin{eqnarray}\label{multiint}
\beta_{\k}(0)&=&\sum_{m=0}^\infty\int_{-\infty}^{0}d\tau_{0}
\frac{\dot{\omega}(\tau_{0})}{2\omega(\tau_{0})}e^{-2iK(\tau_{0})} \nn\\
&\times&\prod_{n=1}^{m}\int_{-\infty}^{\tau_{n-1}}dt_{n}\frac{\dot{\omega}(t_{n})}{2\omega(t_{n})}e^{2iK(t_{n})}
\int_{-\infty}^{t_{n}}d\tau_{n}\frac{\dot{\omega}(\tau_{n})}{2\omega(\tau_{n})}e^{-2iK(\tau_{n})}.
\end{eqnarray}
Each of these integrals can be calculated using a saddle point approximation.
Those integrals are dominated by the regions around the turning point, i.e., $\omega(\tau_{p}^{\pm})=0$, where the superscript $\pm$ denotes the
two conjugate pairs in the complex plane of $\tau$.
More precisely, by deforming the contour of integration, we consider the singularities for the turning point for which
\begin{equation}\label{impart}
\Im[K(\tau_{p})]<0.
\end{equation}
From now on, the subscript $\pm$ will be dropped, and we will consider $\tau_{p}$ the turning point which corresponds to~(\ref{impart}).
Following \cite{Berry:1982}, it is possible to describe the behavior of $\omega^2(\tau)$ near the turning point assuming first order singularity
which is the case contemplating Eq.~(\ref{omega}),
\begin{equation}\label{model}
\omega^2(\tau)\simeq A(\tau-\tau_{p}),
\end{equation}
with $A$ being a constant which can be calculated. One can find then an expression for $K(\tau)$ near the turning point
\begin{align}
\label{turneq1}
K(\tau)&\simeq K(\tau_{p})+\frac{2}{3}A(\tau-\tau_{p})^{\frac{3}{2}}, \\
\label{turneq2}
\frac{\dot{\omega}(\tau)}{\omega(\tau)}&\simeq\frac{1}{3\big(K(\tau)-K(\tau_{p})\big)}
\frac{dK(\tau)}{d\tau}.
\end{align}
Changing variables to $\xi_{n}=K(\tau_{p})-K(\tau_{n})$ and $\eta_{n}=K(\tau_{p})-K(\tau'_{n})$ one gets an approximate expression for the integrals
\begin{equation}\label{semibeta}
\beta_{\k}(0)\simeq-2i\pi e^{-2iK(\tau_{p})}\sum_{m=0}^{\infty}\frac{(-1)^m}{6^{m+1}}I_{m},
\end{equation}
where
\begin{equation}\label{im}
I_{m}=\frac{1}{2i\pi}\int_{-\infty}^{\infty}d\xi_{0}\frac{e^{i\xi_{0}}}{\xi_{0}}\prod_{n=1}^{m}
\int_{-\infty}^{\xi_{n-1}}d\eta_{n}\frac{e^{-i\eta_{n}}}{\eta_{n}}\int_{\eta_{n}}^{\infty}
d\xi_{n}\frac{e^{i\xi_{n}}}{\xi_{n}}=\frac{\pi^{2m}}{(2m+1)!}.
\end{equation}
The final results read then
\begin{equation}\label{final}
n_{\k}=\left|e^{-2iK(\tau_{p})}\right|^2.
\end{equation}
For the semiclassical approximation to hold, one needs the notion of adiabatic vacuum in the asymptotic future.
Hence, the semiclassical approximation holds if the relation~(\ref{assume}) is satisfied.
The remaining step is to compute the integral~(\ref{keq}). The turning point is given by
\begin{equation}\label{taup}
\tau_{p}=\frac{1}{k}\Big[i\kappa-i\Big(\rho^{2}+\kappa^{2}\Big)^{\frac{1}{2}}\Big],
\end{equation}
where the coefficients $\rho$ and $\kappa$ have been defined in Eqs.~(\ref{lambda}) and~(\ref{kappa}), respectively.
Then one can find the imaginary part of $K(\tau)$
\begin{equation}\label{findim}
\Im[K(\tau_{p})]=-\pi\big(\rho+\lambda r\big)\theta\big(-\lambda r\big),
\end{equation}
where the Heaviside step function $\theta$ is there to ensure that the condition~(\ref{impart}) holds.
Recall our convention $\lambda=-eE/H^{2}$, and hence Eq.~(\ref{findim}) implies that, e.g., a particle with charge $e>0$ is only created with a
momentum $k_{x}>0$. Again, we see that in the semiclassical limit, the upward tunneling is suppressed and only the screening direction or downward
tunneling stays. The number of pairs in the semiclassical limit is eventually given by
\begin{equation}\label{main}
n_{\k}=\exp\Big[-2\pi\Big(\rho+\lambda r\Big)\Big]\theta(-\lambda r).
\end{equation}
The pair creation rate is defined as
\begin{equation}\label{semidif}
\Gamma=\frac{1}{\Delta V}\int\frac{d^{d}k}{(2\pi)^{d}}\,n_{\k},
\end{equation}
where $\Delta V$ is defined by Eq.~(\ref{slice}).
As before, the procedure then is to transform the $k$ integral into a $\tau$ integral by using an estimate for the time when most of the particles
are created; see Eq.~(\ref{extremum}).
Using Eqs.~(\ref{formul}), (\ref{formula}), and~(\ref{area}), it is then possible to present the final expression for the scalar pair creation rate
in $\dsd$ under the influence of a constant electric field,
\begin{equation}\label{smifinal}
\Gamma=\frac{H^{D}}{(2\pi)^{d}}\rho^{d}|\lambda|^{\frac{1-d}{2}}e^{-2\pi(\rho-|\lambda|)}.
\end{equation}
A common feature regardless of the number of spatial dimensions and of the bosonic or fermionic nature of the particle is that the physical number
density defined in Eq.~(\ref{density}) is constant with respect to conformal time.
It signals that pair creation in $\dsd$ from electric and gravitational fields exactly balances the dilution from the expansion of the universe.
This implies that the population of scalars is always dominated by the particle created within a Hubble time \cite{Kobayashi:2014zza}.
This observation is important when it comes to study the backreaction to the electric field \cite{Stahl:2016geq}.
It is interesting to note that when one changes the space dimension, what changes is the prefactor before the exponential.
Indeed, the semiclassical approximation is an expansion in $\hbar$ to first order.
The exponential factor is the classical trajectory which is not a function of the dimension $d$.
However, the one loop integration depends on $d$, and hence the prefactor to the classical trajectory is a function $d$.
Before concluding, we should remark that under the semiclassical condition~(\ref{assume}) the result~(\ref{rate}) obtained using standard methods
reduced to the semiclassical result~(\ref{smifinal}).
Therefore, the flat spacetime limit of the semiclassical pair creation rate~(\ref{smifinal}) is equal to the result presented in Eq.~(\ref{limit}).
%%%%%%%%%%%%%%%%%%%%%%%%%%%%%%%%%%%%%%%%%%%%%%%%%%%%%%%%%%%%%%%%%%%%%%%%%%%%%%%%%%%%%%%%%%%%%%%%%%%%%%%%%%%%%%%%%%%%%%%%%%%%%%%%%%%%%%%%%%%%%%%%%%%%%%%%%%%%%%%
%%%%%%%%%%%%%%%%%%%%%%%%%%%%%%%%%%%%%%%%%%%%%%%%%%%%%%%%%%%%%%%%%%%%%%%%%%%%%%%%%%%%%%%%%%%%%%%%%%%%%%%%%%%%%%%%%%%%%%%%%%%%%%%%%%%%%%%%%%%%%%%%%%%%%%%%%%%%%%%
\section{\label{app:math}useful mathematical functions}
In this appendix, we have represented some useful relations and properties of mathematical functions needed in this article.
More relations can be found in, e.g., \cite{book:Nist}.
%%%%%%%%%%%%%%%%%%%%%%%%%%%%%%%%%%%%%%%%%%%%%%%%%%%%%%%%%%%%%%%%%%%%%%%%%%%%%%%%%%%%%%%%%%%%%%%%%%%%%%%%%%%%%%%%%%%%%%%%%%%%%%%%%%%%%%%%%%%%%%%%%%%%%%%%%%%%%%%
%%%%%%%%%%%%%%%%%%%%%%%%%%%%%%%%%%%%%%%%%%%%%%%%%%%%%%%%%%%%%%%%%%%%%%%%%%%%%%%%%%%%%%%%%%%%%%%%%%%%%%%%%%%%%%%%%%%%%%%%%%%%%%%%%%%%%%%%%%%%%%%%%%%%%%%%%%%%%%%
\subsection{\label{app:aw}Whittaker functions}
The Whittaker differential equation is
\begin{equation}\label{whittaker}
\frac{d^{2}}{dz^{2}}F(z)+\Big(-\frac{1}{4}+\frac{\kappa}{z}+\frac{\frac{1}{4}-\gamma^{2}}{z^{2}}\Big)F(z)=0.
\end{equation}
It has the two linearly independent solutions, namely, $\wwp(z)$ and $\wmp(z)$. The needed connection formulas are
\begin{eqnarray}
\label{conectionw}
\wwp(z)&=&\wwm(z), \\
\label{conectionm}
\wmp(e^{\pm i\pi}z)&=&\pm ie^{\pm\gamma i\pi}\m_{-\kappa,\gamma}(z).
\end{eqnarray}
The asymptotical expansion of the Whittaker functions as $|z|\rightarrow\infty$ are given by
\begin{align}
\label{win}
\wwp(z)&\sim e^{-\frac{z}{2}}z^{\kappa}, \\
\wmp(z)&\sim\frac{\Gamma(1+2\gamma)}{\Gamma(\frac{1}{2}+\gamma-\kappa)}
\,e^{\frac{z}{2}}z^{-\kappa} +\frac{\Gamma(1+2\gamma)}{\Gamma(\frac{1}{2}+\gamma+\kappa)}
 \,e^{-\frac{z}{2}\pm(\frac{1}{2}+\gamma-\kappa)\pi i}z^{\kappa}, \nn\\
\label{min}
-\frac{1}{2}\pi+\delta&\leq\pm{\rm ph}(z)\leq\frac{3}{2}\pi-\delta,
\end{align}
where $\delta$ is an arbitrary small positive constant.
In the limit $|z|\rightarrow0$, the asymptotically expansions are given by
\begin{align}
\label{mout}
&\wmp(z)\sim z^{\frac{1}{2}+\gamma}, \\
\label{wout}
&\wwp(z)\sim\frac{\Gamma(2\gamma)}{\Gamma(\frac{1}{2}+\gamma-\kappa)}
z^{\frac{1}{2}-\gamma}+\frac{\Gamma(-2\gamma)}{\Gamma(\frac{1}{2}-\gamma-\kappa)}
z^{\frac{1}{2}+\gamma}, & 0\leq\Re(\gamma)<\frac{1}{2}, &&\gamma\neq0.
\end{align}
Finally, some useful Wronskians are
\begin{eqnarray}
\label{wrww}
{\cal{W}}\Big\{\wwp(z),\w_{-\kappa,\gamma}(e^{\pm i\pi}z)\Big\}&=&e^{\mp i\pi\kappa}, \\ \label{wrmm}
{\cal{W}}\Big\{\wmp(z),\m_{\kappa,-\gamma}(z)\Big\}&=&-2\gamma, \\ \label{wrmw}
{\cal{W}}\Big\{\wwp(z),\wmp(z)\Big\}&=&\frac{\Gamma(1+2\gamma)}{\Gamma(\frac{1}{2}+\gamma-\kappa)}.
\end{eqnarray}
%%%%%%%%%%%%%%%%%%%%%%%%%%%%%%%%%%%%%%%%%%%%%%%%%%%%%%%%%%%%%%%%%%%%%%%%%%%%%%%%%%%%%%%%%%%%%%%%%%%%%%%%%%%%%%%%%%%%%%%%%%%%%%%%%%%%%%%%%%%%%%%%%%%%%%%%%%%%%%%
%%%%%%%%%%%%%%%%%%%%%%%%%%%%%%%%%%%%%%%%%%%%%%%%%%%%%%%%%%%%%%%%%%%%%%%%%%%%%%%%%%%%%%%%%%%%%%%%%%%%%%%%%%%%%%%%%%%%%%%%%%%%%%%%%%%%%%%%%%%%%%%%%%%%%%%%%%%%%%%
\subsection{\label{app:bessel}Modified Bessel functions}
The modified Bessel function has integral representation along the real line
\begin{equation}\label{bessel}
\I_{\nu}(z)=\frac{z^{\nu}}{2^{\nu}\pi^{\frac{1}{2}}\Gamma(\nu+\frac{1}{2})}
\int_{0}^{\pi}(\sin\theta)^{2\nu}e^{\pm z\cos\theta}d\theta.
\end{equation}
If $n$ is an integer, then
\begin{equation}\label{parity}
\I_{\nu}(e^{n\pi i}z)=e^{n\nu\pi i}\I_{\nu}(z).
\end{equation}
When $\nu$ is fixed and $z\rightarrow 0$,
\begin{align}\label{origin}
\I_{\nu}(z)&\sim\frac{z^{\nu}}{2^{\nu}\Gamma(1+\nu)}, & \nu&\neq-1,-2,-3,\ldots.
\end{align}
When $\nu$ is fixed and $z\rightarrow\infty$,
\begin{align}\label{infty}
\I_{\nu}(z)&\sim\frac{e^{z}}{\sqrt{2\pi z}}, & \big|\mathrm{ph}(z)\big|&\leq\frac{\pi}{2}-\delta.
\end{align}
In the cases $\nu=-\frac{1}{2}$ and $\nu=\frac{1}{2}$, the relations
\begin{eqnarray}
\I_{-\frac{1}{2}}(z)&=&\sqrt{\frac{2}{\pi z}}\cosh(z), \label{cosh} \\
\I_{\frac{1}{2}}(z)&=&\sqrt{\frac{2}{\pi z}}\sinh(z), \label{sinh}
\end{eqnarray}
are satisfied. The following mathematical formulas can be shown:
\begin{eqnarray}
\int_{0}^{\pi}(\sin\theta)^{2\nu}d\theta&=&\frac{\sqrt{\pi}}{\Gamma\big(1+\nu\big)}
\Gamma\big(\frac{1}{2}+\nu\big), \label{formul} \\
\lim_{|\lambda|\rightarrow\infty}\int_{\frac{\pi}{2}}^{\pi}(\sin\theta)^{\nu}
e^{-2\pi|\lambda|\cos\theta}d\theta&=&
\frac{\Gamma\big(\frac{\nu+1}{2}\big)}{2(\pi\lambda)^{\frac{\nu+1}{2}}}e^{2\pi|\lambda|}. \label{formula}
\end{eqnarray}
%%%%%%%%%%%%%%%%%%%%%%%%%%%%%%%%%%%%%%%%%%%%%%%%%%%%%%%%%%%%%%%%%%%%%%%%%%%%%%%%%%%%%%%%%%%%%%%%%%%%%%%%%%%%%%%%%%%%%%%%%%%%%%%%%%%%%%%%%%%%%%%%%%%%%%%%%%%%%%%
%%%%%%%%%%%%%%%%%%%%%%%%%%%%%%%%%%%%%%%%%%%%%%%%%%%%%%%%%%%%%%%%%%%%%%%%%%%%%%%%%%%%%%%%%%%%%%%%%%%%%%%%%%%%%%%%%%%%%%%%%%%%%%%%%%%%%%%%%%%%%%%%%%%%%%%%%%%%%%%
\subsection{\label{app:spher}Spherical coordinates}
In order to evaluate the integrals~(\ref{intk}) and~(\ref{semidif}), we make use of the spherical coordinates to decompose the momentum vector $\k$
in the flat $d$-dimensional Euclidean space. Hence, in this space the volume element is
\begin{equation}\label{velement}
d^{d}k=d\Sigma_{d-1}k^{d-1}dk,
\end{equation}
where $d\Sigma_{d-1}$ is the area element of the unit sphere in the $d$-dimensional Euclidean space.
Convenient coordinates on this sphere are specified by
\begin{eqnarray}\label{cordinat}
\omega^{1}&=&\cos\theta_{1}, \nn\\
\omega^{2}&=&\sin\theta_{1}\cos\theta_{2}, \nn\\
&\vdots& \nn\\
\omega^{d-1}&=&\sin\theta_{1}\cdots\sin\theta_{d-2}\cos\theta_{d-1}, \nn\\
\omega^{d}&=&\sin\theta_{1}\cdots\sin\theta_{d-2}\sin\theta_{d-1},
\end{eqnarray}
where $0\leq\theta_{i}<\pi$ for $1\leq i\leq d-2$ and $0\leq\theta_{d-1}<2\pi$.
Then, the metric on the sphere is
\begin{equation}\label{sphere}
d\varpi_{d-1}^{2}=\sum_{i=1}^{d}(d\omega^{i})^{2}=d\theta_{1}^{2}+\sin^{2}\theta_{1}d\theta_{2}^{2}+\cdots
+\sin^{2}\theta_{1}\cdots\sin^{2}\theta_{d-2}d\theta_{d-1}^{2},
\end{equation}
and consequently, the area element is
\begin{equation}\label{aelement}
d\Sigma_{d-1}=(\sin\theta_{1})^{d-2}\cdots\sin\theta_{d-2}d\theta_{1}\cdots d\theta_{d-1}.
\end{equation}
Therefore the area of the sphere is
\begin{equation}\label{area}
\int d\Sigma_{d-1}=\frac{2\pi^{\frac{d}{2}}}{\Gamma(\frac{d}{2})}.
\end{equation}
Using Eqs.~(\ref{bessel}), (\ref{formul}), (\ref{aelement}), and~(\ref{area}) the following formula can be shown:
\begin{equation}\label{areaint}
\int d\Sigma_{d-1}e^{2\pi\lambda\cos\theta_{1}}=2\pi\lambda^{1-\frac{d}{2}}\I_{\frac{d}{2}-1}(2\pi\lambda).
\end{equation}
%%%%%%%%%%%%%%%%%%%%%%%%%%%%%%%%%%%%%%%%%%%%%%%%%%%%%%%%%%%%%%%%%%%%%%%%%%%%%%%%%%%%%%%%%%%%%%%%%%%%%%%%%%%%%%%%%%%%%%%%%%%%%%%%%%%%%%%%%%%%%%%%%%%%%%%%%%%%%%%
%%%%%%%%%%%%%%%%%%%%%%%%%%%%%%%%%%%%%%%%%%%%%%%%%%%%%%%%%%%%%%%%%%%%%%%%%%%%%%%%%%%%%%%%%%%%%%%%%%%%%%%%%%%%%%%%%%%%%%%%%%%%%%%%%%%%%%%%%%%%%%%%%%%%%%%%%%%%%%%
\section{\label{app:int}Computation of the integral for the current}
In this appendix, the computation of the current integral~(\ref{curent}) is reviewed.
We follow the same integration procedure as performed in \cite{Frob:2014zka} for a one-dimensional and \cite{Kobayashi:2014zza} for a
three-dimensional momentum integral. We deal with the following integral:
\begin{equation}\label{integral}
\mathcal{J}:=\lim_{\Lambda\rightarrow\infty}\int_{-1}^{1}\frac{dr}{\sqrt{1-r^{2}}}\int_{0}^{\Lambda}dp
\big(rp-\lambda\big)e^{\lambda r\pi}\big|\w_{-i\lambda r,\gamma}(-2ip)\big|^{2}.
\end{equation}
We will use the Mellin-Barnes representation of the Whittaker function
\begin{eqnarray}\label{mellin}
\wwp(z)&=&e^{-\frac{z}{2}}\int_{-i\infty}^{+i\infty}\frac{ds}{2\pi i}
\frac{\Gamma(\frac{1}{2}+\gamma+s)\Gamma(\frac{1}{2}-\gamma+s)\Gamma(-\kappa-s)}
{\Gamma(\frac{1}{2}+\gamma-\kappa)\Gamma(\frac{1}{2}-\gamma-\kappa)}z^{-s}, \nn\\
\big|{\rm ph}(z)\big|&<&\frac{3\pi}{2}, \hspace{1cm} \frac{1}{2}\pm\gamma-\kappa\neq0,-1,-2,\ldots,
\end{eqnarray}
where the contour of integration separates the poles of $\Gamma(\frac{1}{2}+\gamma+s)\Gamma(\frac{1}{2}-\gamma+s)$ from those of $\Gamma(-\kappa-s)$ \cite{book:Nist}.
Based on the definition~(\ref{gamma}), it depends on the range of the involved parameters whether $\gamma$ is real or purely imaginary.
However, by virtue of the relation~(\ref{conectionw}), in the integral~(\ref{integral}) we have
$\big(\w_{-i\lambda r,\gamma}(-2ip)\big)^{\ast}=\w_{i\lambda r,\gamma}(2ip)$.
Then, the integral~(\ref{integral}) can be rewritten
\begin{eqnarray}\label{contourint}
\mathcal{J}&=&\lim_{\Lambda\rightarrow\infty}\int_{-1}^{1}\frac{dr}{\sqrt{1-r^{2}}}c_{r}
\int_{-i\infty}^{+i\infty}\frac{ds}{2\pi i}\Gamma(\frac{1}{2}+\gamma+s)\Gamma(\frac{1}{2}-\gamma+s)
\Gamma(i\lambda r-s) \nn\\
&\times&\int_{-i\infty}^{+i\infty}\frac{dt}{2\pi i}\Gamma(\frac{1}{2}+\gamma+t)\Gamma(\frac{1}{2}-\gamma+t)
\Gamma(-i\lambda r-t)e^{\frac{i\pi}{2}(s-t)}2^{-s-t} \nn\\
&\times&\int_{0}^{\Lambda}dp\big(rp-\lambda\big)p^{-s-t},
\end{eqnarray}
where $c_{r}$ is defined as
\begin{equation}\label{cr}
c_{r}=e^{\pi\lambda r}\Big(\Gamma\big(\frac{1}{2}+\gamma+i\lambda r\big)
\Gamma\big(\frac{1}{2}-\gamma+i\lambda r\big)\Gamma\big(\frac{1}{2}+\gamma-i\lambda r\big)
\Gamma\big(\frac{1}{2}-\gamma-i\lambda r\big)\Big)^{-1}.
\end{equation}
If we choose both $s$ and $t$ integration contours to run in a similar way as Ref.~\cite{Kobayashi:2014zza}, then we obtain the final result
\begin{eqnarray}\label{computed}
\mathcal{J}&=&-\frac{\pi}{2}\lambda\lim_{\Lambda\rightarrow\infty}\Lambda
+\frac{\pi}{4}\lambda\gamma\cot(2\pi\gamma)
+\frac{\gamma}{4\sin(2\pi\gamma)}\Big(3\I_{1}(2\pi\lambda)-2\pi\lambda\I_{0}(2\pi\lambda)\Big) \nn\\
&+&\frac{i}{2\sin(2\pi\gamma)}\int_{-1}^{1}\frac{dr}{\sqrt{1-r^{2}}}b_{r}
\Big\{\big(e^{2\pi\lambda r}+e^{-2\pi i\gamma}\big)\psi\big(\frac{1}{2}+i\lambda r-\gamma\big) \nn\\
&-&\big(e^{2\pi\lambda r}+e^{2\pi i\gamma}\big)\psi\big(\frac{1}{2}+i\lambda r+\gamma\big)\Big\},
\end{eqnarray}
where $b_{r}$ is given by Eq.~(\ref{br}).
%%%%%%%%%%%%%%%%%%%%%%%%%%%%%%%%%%%%%%%%%%%%%%%%%%%%%%%%%%%%%%%%%%%%%%%%%%%%%%%%%%%%%%%%%%%%%%%%%%%%%%%%%%%%%%%%%%%%%%%%%%%%%%%%%%%%%%%%%%%%%%%%%%%%%%%%%%%%%%%
%%%%%%%%%%%%%%%%%%%%%%%%%%%%%%%%%%%%%%%%%%%%%%%%%%%%%%%%%%%%%%%%%%%%%%%%%%%%%%%%%%%%%%%%%%%%%%%%%%%%%%%%%%%%%%%%%%%%%%%%%%%%%%%%%%%%%%%%%%%%%%%%%%%%%%%%%%%%%%%

%%%%%%%%%%%%%%%%%%%%%%%%%%%%%%%%%%%%%%%%%%%%%%%%%%%%%%%%%%%%%%%%%%%%%%%%%%%%%%%%%%%%%%%%%%%%%%%%%%%%%%%%%%%%%%%%%%%%%%%%%%%%%%%%%%%%%%%%%%%%%%%%%%%%%%%%%%%%%%%
%%%%%%%%%%%%%%%%%%%%%%%%%%%%%%%%%%%%%%%%%%%%%%%%%%%%%%%%%%%%%%%%%%%%%%%%%%%%%%%%%%%%%%%%%%%%%%%%%%%%%%%%%%%%%%%%%%%%%%%%%%%%%%%%%%%%%%%%%%%%%%%%%%%%%%%%%%%%%%%
\end{document}